\pdfoutput=1
\documentclass[11pt]{article}

\usepackage[]{acl}

\usepackage{times}
\usepackage{latexsym}

\usepackage[T1]{fontenc}
\usepackage[utf8]{inputenc}

\usepackage{microtype}
\usepackage{listings}
\usepackage{subfig}
\usepackage{booktabs} % for professional tables
\usepackage{amsmath}
\usepackage{amsthm}
\usepackage{amsfonts}       % blackboard math symbols
\usepackage{nicefrac}       % compact symbols for 1/2, etc.
\usepackage{xspace}
\usepackage{xcolor}
\usepackage{backnaur}
\usepackage{multirow}
\usepackage{makecell}
\usepackage{appendix}
\usepackage{enumitem}
\usepackage{circledsteps}
\usepackage{color}
\usepackage{colortbl}
\usepackage{graphicx}
\usepackage{fixltx2e}
\usepackage{hyperref}

\theoremstyle{definition}
\newtheorem{definition}{Definition}

\newcommand{\refequ}[1]{Equation~(\ref{#1})}
\newcommand{\reffig}[1]{Figure~\ref{#1}}
\newcommand{\refsec}[1]{\S\ref{#1}} % \textsection
\newcommand{\reftab}[1]{Table~\ref{#1}}

\newcommand{\tabincell}[2]{\begin{tabular}{@{}#1@{}}#2\end{tabular}}

\def\eg{\textit{e.g.}\xspace}
\def\Eg{\textit{E.g.}\xspace}

\def\etc{\textit{etc.}\xspace}
\def\ie{\textit{i.e.}\xspace}

\usepackage{enumitem}
\setlist[itemize]{align=parleft,left=0pt..1em}
\setenumerate[1]{itemsep=0pt,partopsep=0pt,parsep=\parskip,topsep=5pt}
\setitemize[1]{itemsep=0pt,partopsep=0pt,parsep=\parskip,topsep=5pt}
\setdescription{itemsep=0pt,partopsep=0pt,parsep=\parskip,topsep=5pt}

\title{AnaMeta: A Table Understanding Dataset of Field Metadata Knowledge Shared by Multi-dimensional Data Analysis Tasks}

\makeatletter
\newcommand{\printfnsymbol}[1]{%
  \textsuperscript{\@fnsymbol{#1}}%
} 
\makeatother

\author{
Xinyi He\textsuperscript{\rm 1}\thanks{\indent The contributions by Xinyi He, Mingjie Zhou, Jialiang Xu, Tianle Li and Yijia Shao have been conducted and completed during their internships at Microsoft Research Asia, Beijing, China.}\hspace{0.5em}
Mengyu Zhou\textsuperscript{\rm 2}\thanks{\indent Corresponding author.}\hspace{0.5em}
Mingjie Zhou\textsuperscript{\rm 3}\printfnsymbol{1}\hspace{0.5em}
Jialiang Xu\textsuperscript{\rm 4}\printfnsymbol{1}\hspace{0.5em}
Xiao Lv\textsuperscript{\rm 2}\hspace{0.5em}\\ 
\textbf{Tianle Li}\textsuperscript{\rm 5}\printfnsymbol{1}\hspace{0.5em}
\textbf{Yijia Shao}\textsuperscript{\rm 6}\printfnsymbol{1}\hspace{0.5em}
\textbf{Shi Han}\textsuperscript{\rm 2}\hspace{0.5em}
\textbf{Zejian Yuan}\textsuperscript{\rm 1}\hspace{0.5em}
\textbf{Dongmei Zhang}\textsuperscript{\rm 2}\hspace{0.5em} \\
\textsuperscript{\rm 1} Xi'an Jiaotong University
\textsuperscript{\rm 2} Microsoft Research 
 \textsuperscript{\rm 3} The University of Hong Kong \\
\textsuperscript{\rm 4} University of Illinois at Urbana-Champaign
\textsuperscript{\rm 5} University of Waterloo
\textsuperscript{\rm 6} Peking University \\
\texttt{\href{mailto:hxyhxy@stu.xjtu.edu.cn}{hxyhxy@stu.xjtu.edu.cn}},
\texttt{\href{mailto:mjzhou@connect.hku.hk}{mjzhou@connect.hku.hk}},
\texttt{\href{mailto:jx17@illinois.edu}{jx17@illinois.edu}}, \\ 
\texttt{\href{t29li@uwaterloo.ca}{t29li@uwaterloo.ca}},
\texttt{\href{shaoyj@pku.edu.cn}{shaoyj@pku.edu.cn}},
\texttt{\href{yuan.ze.jian@xjtu.edu.cn}{yuan.ze.jian@xjtu.edu.cn}},\\
\texttt{\{\href{mailto:mezho@microsoft.com}{mezho}, \href{mailto:xilv@microsoft.com}{xilv}, \href{mailto:shihan@microsoft.com}{shihan}, \href{mailto:dongmeiz@microsoft.com}{dongmeiz}\}@microsoft.com}}

\begin{document}
\maketitle
\begin{abstract}
    % Many data analysis tasks heavily rely on a deep understanding of tables (multi-dimensional data). A few previous works have noticed that such analysis knowledge (field metadata) of table fields is shared across the tasks. To represent field metadata, we propose AnaMeta benchmark, including tasks and corpus. We identify four AnaMeta tasks: Measure/dimension dichotomy, common field roles, semantic field type, and default aggregation function. While those AnaMeta face challenges of insufficient supervision signals. A large corpus of AnaMeta collected from spreadsheet, web table datasets and Synthetic dataset. We use smart supervision from down stream tasks, public dataset, and manual labels.

% To inference and applicate AnaMeta, we propose AnaMeta Framework including model to represent AnaMeta and 4 interfaces to applicate it to downstream applications. Our multi-tasking model fuses field distribution and knowledge graph information into pre-trained tabular models. we propose the use of four different interfaces for injecting knowledge: directly injecting metadata ID, using metadata output embeddings, using metadata output sentences, and using metadata tasks as pre-training tasks.

% We conduct a series of experiences to evaluate the effective of AnaMeta Framework. Our best model outperforms a series of baselines that are based on rules, traditional machine learning methods, and pre-trained tabular models. Meanwhile, experiences have shown that AnaMeta with different interfaces promotes downstream applications, such as chart recommendation, and natural language QA...

Tabular data analysis is performed every day across various domains. It requires an accurate understanding of field semantics to correctly operate on table fields and find common patterns in daily analysis. In this paper, we introduce the AnaMeta dataset, a collection of 467k tables with derived supervision labels for four types of commonly used field metadata: measure/dimension dichotomy, common field roles, semantic field type, and default aggregation function. We evaluate a wide range of models for inferring metadata as the benchmark. We also propose a multi-encoder framework, called KDF, which improves the metadata understanding capability of tabular models by incorporating distribution and knowledge information. Furthermore, we propose four interfaces for incorporating field metadata into downstream analysis tasks.
\end{abstract}

\section{Introduction}
\label{sec:intro}

\begin{figure*}
    \centering
   \includegraphics[width=2 \columnwidth]{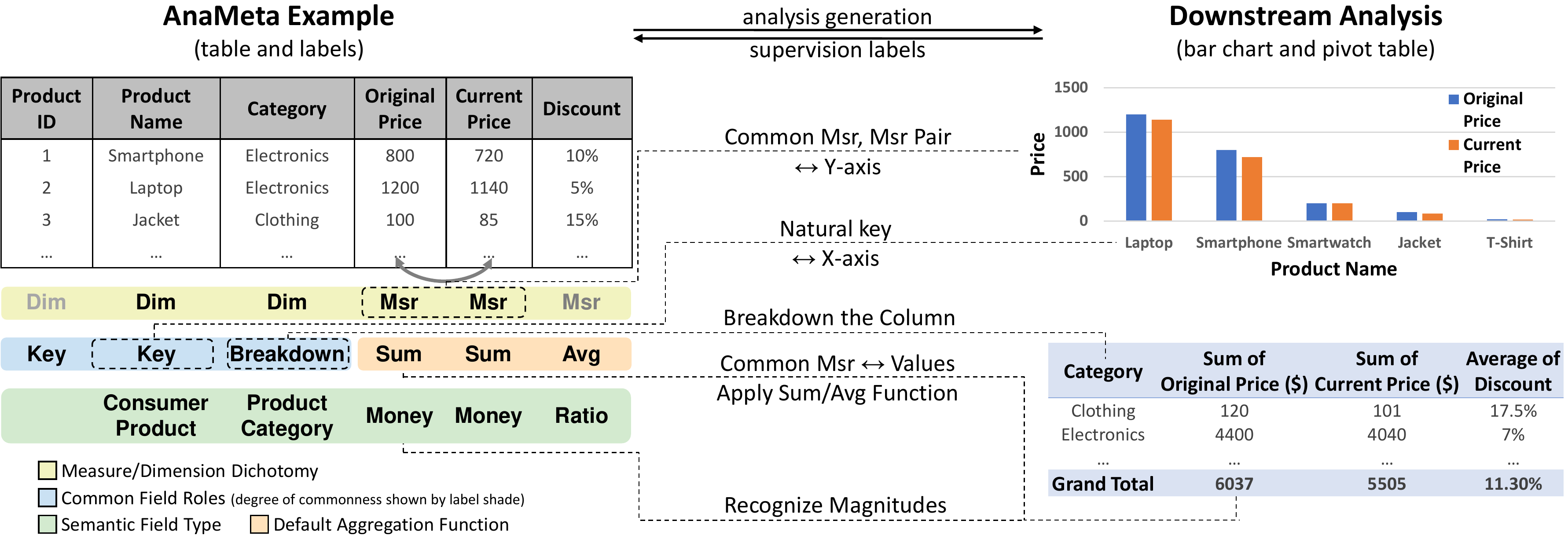}
    \caption{An Example Table of AnaMeta dataset with Field Metadata Task Supervision and Downstream Analysis.}
    \label{fig:overall}
    \vspace{-3mm}
\end{figure*}

Tabular data analysis is performed every day in popular tools such as Excel, Google Sheets, and Tableau~\cite{rebman2022industry} for a wide range of domains including education, research, engineering, finance, HR, \etc To help non-expert users, various machine learning tasks are proposed to automate and accelerate the analysis process. For example, TableQA \& Text2SQL~\cite{dong2016nl2sql,km2021survey}, analysis \& visualization recommendations ~\cite{zhou2021charts,aoyu2021survey}, insights mining~\cite{ding2019quickinsights,law2020vis}, \etc

These analysis tasks require an accurate understanding of field semantics to correctly operate on table fields (or columns) and to further find common patterns in daily analysis. Such \textit{analysis knowledge} of field semantics is often shared across multiple tasks. In real-world applications, we also call it \textbf{field metadata} in contrast to the raw tabular input which does not directly provide this information.

For example, \textbf{measure} / \textbf{dimension} dichotomy is one such metadata used in Tableau~\cite{tableau201849} and Excel~\cite{ding2019quickinsights} across diverse features. Its definition is inspired by dimensional modeling in databases~\cite{golfarelli1998dimensional,kimball2013data}. 
It involves categorizing each field in a table as either measure or dimension. 
A measure (field) contains numerical measurement results on which calculations can be made, such as sum, count, average, minimum, and maximum. \Eg, ``Price'' and ``Discount'' fields of \reftab{fig:overall} are measures.
On the other hand, a dimension (field) contains categorical information and can be used for filtering, grouping, and labeling. \Eg, ``Product Name'' and ``Category'' fields of \reftab{fig:overall} are dimensions.
Infeasible analysis might be generated based on incorrect classification of measures and dimensions in a table, because the feasible operations for measure and dimension are largely different. \Eg, for \reftab{fig:overall}, it is impossible to draw a bar chart without measures -- only mapping dimension ``Product Name'' to x-axis and dimension ``Product Id'' to y-axis.

Beyond the simple dichotomy of measure / dimension, what are other types of useful field metadata? For each type of metadata, where do we find labels to evaluate or train models? To address these questions, in this paper we define the following four types of commonly used field metadata, and collect the \textbf{AnaMeta} (\textbf{Ana}lysis \textbf{Meta}data) dataset incorporating tables from 3 diverse sources with derived supervision labels of the metadata:

\refsec{sec:msr-dim} \textit{Measure/dimension dichotomy}: Categorizing each field in a table as a measure or dimension.

\refsec{sec:common-msr-dim} \textit{Common field roles}: Identifying whether a measure field is commonly used as an analysis target, whether a dimension field is a natural key or commonly used for breakdown operation.

\refsec{sec:msr-pair-type} \textit{Semantic field type}: 
Identifying types of measure fields and dimension fields from our established measure type taxonomy 
(\eg, ``Money'', ``Ratio'' in \reffig{fig:overall}) 
and a dimension type taxonomy provided by a knowledge graph (\eg, ``Consumer Product'' in \reffig{fig:overall}), 
and determining whether two measures %from a table 
are comparable with each other.

\refsec{sec:default-agg} \textit{Default aggregation function}: 
Identifying the most appropriate default aggregation function for a measure field. 
(\eg, ``Avg'', ``Sum''in \reffig{fig:overall})

The collection of the AnaMeta dataset is a complex process that requires a significant amount of data and supervision. Due to the limitations of obtaining this information directly from raw tables, we have employed a multi-source approach to collecting AnaMeta, which includes spreadsheet datasets (45,361 tables), public web table datasets (404,152 tables), and public synthetic datasets (17,995 tables).
Supervision is smartly collected from downstream tasks, manual labels through crowdsourcing, and existing information attached to tables(\refsec{sec:supervision}). 
After preprocessing, AnaMeta contains 152,092 fields with measure/dimension, 149,197 with common field roles, 1,730,494 with semantic field type and 38,030 with aggregation.
Additionally, we have conducted quality inspection and further checks to ensure accuracy and reliability.

Based on the AnaMeta dataset, in the rest of the paper, we first evaluate a wide range of models on inferring metadata in order to check how well they learned the knowledge of field semantics. In \refsec{sec:benchmark}, pre-trained tabular (TURL, TAPAS, TABBIE) and large language (GPT-3/3.5 family) models are compared through additional classification heads and zero- / few-shot prompts. %Also, we compare the performance gap between the traditional ML models (heuristic rules, GBDT, and random forest) with these pre-trained NN models. 
Semantic information captured by the pre-trained tabular models brings great gain to metadata tasks.

To improve the metadata understanding capability of tabular models, in \refsec{sec:method} we further propose a multi-encoder \textbf{KDF} (Knowledge and Distribution Fusion) framework for transformer-based pre-trained tabular models.
\textit{Knowledge fusion} incorporates knowledge graph information (such as lined entities, column type, and properties). \textit{Distribution fusion} adds distribution information by calculating field data statistics.
%The framework adds distribution information by calculating field data statistics and incorporating them through a process called \textit{Distribution Fusion}. Also, knowledge graph information (such as entity linking, column type, and properties) is extracted for each table and incorporated through \textit{Knowledge Fusion}. This allows for a more comprehensive understanding of the data and improves the overall performance of the tabular model. 
KDF outperforms the best baseline by 3.13\% (see \refsec{sec:main_results}). This indicates that successful knowledge and distribution fusion brings a better representation of fields. 

Finally, we demonstrate one general approach to leveraging metadata to enhance downstream analysis tasks -- taking field metadata as an intermediate step and injecting the knowledge through different interfaces for downstream models. 
In order to incorporate metadata into downstream analysis tasks at various stages of the process, four interfaces are proposed in \refsec{sec:interfaces}. %These interfaces include metadata IDs, metadata embeddings, metadata sentences, and metadata pre-training. 
These interfaces provide different forms of metadata, including field metadata classification results, column embeddings, metadata strings of the field, and field metadata tasks as pre-training objectives. 
In \refsec{sec:exp-downstreamtask}, we apply these interfaces on downstream tasks (TableQA and visualization recommendation) and observe that metadata knowledge helps downstream tasks when injected through appropriate interfaces.

In summary, our main contributions are:
\vspace{-2mm}
\begin{itemize}
    \item We collect a large AnaMeta dataset with the definition and taxonomy of field metadata and smart supervision. The dataset and code will be open-sourced in \url{https://github.com/microsoft/AnaMeta}.
    \item A wide range of models are compared on metadata learning and retaining capabilities.
    \item We propose a KDF framework with distribution fusion and knowledge fusion, which bring a better representation of fields.
    \item Four interfaces are proposed to improve the performance of downstream applications at various stages of the process.
\end{itemize}

% End within 2 pages

\section{Metadata Definition}
\label{sec:problem}

In this section, we delve deeper into the concept of metadata and its definitions. To facilitate understanding, we provide an example in \reffig{fig:overall}. When creating charts and pivot tables, individuals typically begin by analyzing common fields and assigning different roles to them, \eg, using the primary key -- "Product Name" field as the x-axis in a bar chart. After selecting the fields, further analysis is required for each field. In the case of pivot tables, it is necessary to determine that the "original prices" field should be summed. We formulate the metadata tasks as machine learning classification tasks with the table as input. The specifics of this formulation can be found in \refsec{sec:app-formulation}.

\subsection{Measure / Dimension Dichotomy}
\label{sec:msr-dim}
Measure and dimension fields play different roles in data analysis. The measure/dimension dichotomy metadata can tell downstream tasks the legal analysis operations for each field, thus could help greatly instruct their search spaces. They are simply defined in \cite{ding2019quickinsights}.

\begin{definition}[Measure]
A \textbf{measure} (MSR) field contains numerical measurement values on which calculations can be made.
\end{definition}
\vspace{-1mm}

\begin{definition}[Dimension]
A \textbf{dimension} (DIM) field contains categorical values. It provides functions of filtering, grouping, and labeling. A dimension is called a \textbf{breakdown} (or group-by) dimension when its values have duplication. Otherwise, a dimension with unique values is a \textbf{key} dimension.
\end{definition}

\subsection{Common Field Roles}
\label{sec:common-msr-dim}

Fields have been identified with measure and dimension, while not all of them are highly regarded. In daily analysis activities, measures and dimensions with some semantic meanings are more frequently selected. In other words, there are common patterns about which fields are more preferred than others within a table. As shown in \reffig{fig:overall}, we mark common preferences of measure, breakdown dimensions, and key dimensions by shade (darker means more preferred).

\begin{definition}[Natural Key]
\textbf{Natural Key} is a dimension field with all unique data values and uses them to represent each record in semantic terms. 
\end{definition}

% For example, when composing charts from \reftab{tab:A}, because the ``Name'' field is more human-comprehensible, it is more commonly chosen than ``Student ID'' as a key dimension mapping to the x-axis.

\begin{definition}[Common Breakdown]
\textbf{Common Breakdown} is the dimension field(s) that are the most commonly used for breaking down (grouping by) among a given table in data analysis. 
\end{definition}
% \vspace{-1mm}
% For example, when selecting group-by (breakdown) dimensions during insights mining in \reftab{tab:A}, ``Department'' has richer semantics and better readability compared to ``Class''. In \reftab{tab:B}, ``Region'' and ``SalesRep'' are commonly grouped by because of their low cardinality and time-invariant qualities, leading to high-quality insights.

\begin{definition}[Common Measure]
\textbf{Common Measure} is the measure field(s) that are the most commonly used for further analysis (\eg, applying aggregation function, composing chart) among a given table in data analysis.
\end{definition}
% \vspace{-1mm}
% For example, when choosing measure from \reftab{tab:B}, most people would analyze ``Sales'' first as it's a more important KPI that a store would care about prior to the number of products sold. 

\subsection{Semantic Field Type}
\label{sec:msr-pair-type}

Semantic types for common fields are important to data cleaning, table interpretation, data analysis, and so on. There are several existing works focusing on identifying semantic types for columns as described in \refsec{sec:data_profiling}, \eg, column type identification.

However, most works focus on dimension fields, especially the subject columns. Less than 5\% of column types denote measure fields, which are statistics on existing column type datasets (Sherlock~\cite{hulsebos2019sherlock}, TURL~\cite{deng2020TURL}, and Semtab~\cite{jimenez2019semtab,cutrona2020semtab}). Thus, we propose semantic field type, including dimension and measure type.

\begin{definition}[Dimension Type]
 The common semantic types for dimension fields, are based on knowledge graph type of entities. We adopt TURL~\cite{deng2020TURL} column types as dimension type. Details are shown in \refsec{sec:app-dimtype}
\end{definition}
% \vspace{-2mm}

\begin{definition}[Measure Type]
\label{eg:msr-type}
 The common mutually exclusive types for measure fields, are based on unit, magnitudes, and entity property. See the taxonomy in \reftab{tab:msr-taxonomy-small}.
\end{definition}

\begin{table}[htbp]
\small
  \centering
  \caption{Measure Type Taxonomy.}
  \resizebox{\linewidth}{!}{%
    \begin{tabular}{ll||ll}
    \toprule
    Category & Type  & Category & Type \\
    \midrule
    \multirow{6}[2]{*}{Dimensionless} & Count (Amount) & \multirow{9}[7]{*}{Scientific} & Length \\
          & Ratio &       & Area \\
          & Angle &       & Volume(Capacity) \\
          & Score &       & Mass(Weight) \\
          & Rank  &       & Power \\
          & Factor/Coefficient &       & Energy \\
\cmidrule{1-2}    Money &       &       & Pressure \\
\cmidrule{1-2}    Data/file size &       &       & Speed \\
\cmidrule{1-2}    \multirow{2}[2]{*}{Time} & Duration &       & Temperature \\
\cmidrule{3-4}
          & Frequency & Others &  \\
    \bottomrule
    \end{tabular}%
  }
  \label{tab:msr-taxonomy-small}%
  % \vspace{-3mm}
\end{table}%

In \reftab{tab:msr-taxonomy-small} you can find the commonly seen measure types in daily data analysis scenarios. These 19 types (except ``Others'') can be further grouped into 5 categories. 
The taxonomy is based on the units in Wikidata\footnote{\url{https://www.wikidata.org/wiki/Wikidata:Units}}, the International System of Units\footnote{\url{https://www.bipm.org/en/measurement-units}} and DBPedia properties in T2D Golden Standard\footnote{\url{http://webdatacommons.org/webtables/goldstandard.html\#toc2}}. 
As we will discuss in \refsec{sec:supervision}, we only keep the measure types with a number of appearances above the threshold. More details in \refsec{sec:app_msr_type}.

% Semantic field types can be helpful in writing rules, constraints, and templates for data analysis systems. For example, in \reftab{tab:A} and \reftab{tab:B}, measure types are shown in round brackets. When knowing ``Products'' belongs to Count and ``Sales'' belongs to Money, for \reftab{tab:B}, a natural language QA system could suggest default queries by easily filling the templates ``The total (Count) by [a group-by dimension]'' and ``Which record has the highest (Money)?''.

Many tables contain multiple same-typed measures, among which further analysis can be made, implying multi-measure analysis. 

\begin{definition}[Measure Pair]
Within a table, a \textbf{measure pair} is a pair of comparable measures -- they should have the same type of unit (including convertible ones), related semantic meanings, and a similar numerical value range.

% Measure pair can assist in the measure type identification. Besides, when measure type in the table is difficult to identify, the measure pair identification is particularly important. For example, in \reftab{tab:A}, a measure pair is marked by an arrow. ``Midterm Exam'' and ``Final Exam'' in \reftab{tab:A} are a pair of scores (in the Dimensionless category). Such measure pairs are frequently analyzed together. \Eg, the chart in \reffig{fig:chartA}. 

\end{definition}

\subsection{Default Aggregation Function}
\label{sec:default-agg}

% Table generated by Excel2LaTeX from sheet 'Sheet1'
\begin{table*}[htbp]
\small
  \centering
  \caption{Supervision Labels and Statistics from Each Dataset. Each row represents one task, and its labels are obtained from different ways and datasets. The last row (``Statistics'') refers to the number of samples with labels.}
  \resizebox{\textwidth}{!}{%
    \begin{tabular}{c|ccccc|c}
    \toprule
    Supervision & Chart dataset & Pivot dataset  & T2D dataset  & TURL dataset & SemTab dataset & Statistics (field or pair) \\
    \midrule
    Measure & Y-axis field & Value field  & Msr. type field & -     & -  & 110,614\\
    Dimension & \tabincell{c}{X-axis field\\(ex. scatter, line chart)} & Rows and columns field & Primary key field & -     & -  & \ \ 41,478\\
    \midrule
    Com. measure & Chart msr. & Pivot msr.  & -     & -     & - & 110,352 pos, 299,268 neg\\
    Com. breakdown & -     & Non-unique dim.  & -     & -     & - & \ \ 18,815 pos, 117,996 neg\\
    Natural key & Single unique dim. & -     & Primary key & -     & - & \ \ 20,030 pos, \ \ 84,496 neg \\
    \midrule
    Dim. type & -    & -     & -     & \tabincell{c}{Column type\\(cell hyperlinks)} & - &1,235,831\\
    Msr. type & -     & -     &\tabincell{c}{Msr. property\\(manual label)}  & -     & \tabincell{c}{Msr. property\\(from synthetic info.)} & \ \ \ 494,663\\
    Msr. pair & Same chart axis msr. & -    & -     & -     & - & \ \ \ \ \ 33,100\\
    \midrule
    Agg. function & -     & Agg. for value field & -     & -     & - &\ \ \ \ \ 38,030\\
    \midrule
    \midrule
    Statistics (schema \& table) & 21,733 \& 36,461& 7,041 \& 8,900&  702 \& 702 & 403,450 \& 403,450 & 17,995 \& 17,995 & -\\
    \bottomrule
    \end{tabular}%
  }
  % \vspace{-3mm}
  \label{tab:corpus}%
\end{table*}%

\begin{definition}[Default Aggregation Function]
\textbf{Default aggregation function} is an aggregation function that is the most commonly used for applying to the measure field. Popular aggregation (AGG) functions and their statistics are shown in \reffig{fig:agg_funcs}.
\end{definition}
% \vspace{-1mm}

% \begin{example}[Default Aggregation Function]
% \label{eg:agg-func}
% The default AGG function for  ``Midterm Exam'' and ``Final Exam'' in \reftab{tab:A} is average (AVG), and summation (SUM) for ``\#Products'' and ``Sales'' in \reftab{tab:B}.

% The Default AGG function can be directly used to avoid further searching efforts in data analysis. \Eg, there is no need to try AVG and other functions before SUM for the pivot table in \reffig{fig:pivotB}.
% \end{example}
% \vspace{-1mm}

\section{AnaMeta Dataset}
\label{sec:corpora}
\label{sec:corpus}

After defining the field metadata, the next challenge is to locate the supervision labels needed for training and evaluation. 
% As discussed in \refsec{sec:intro}, a major challenge for building machine learning models for analysis metadata inference is where to find supervision labels. 
In this section, we will bring several sources together to prepare the labels.

\subsection{Table Sources}
\label{sec:table_datasets}
\subsubsection{Spreadsheet Tables}
From the public Web, we crawled millions of Excel spreadsheet files with English-language tables in them. Lots of these files contain analysis artifacts created by users. From these spreadsheets, we extract the following datasets:

1) Chart dataset: There are 59,797 charts (\eg, the one in \reffig{fig:overall}) created from 36,461 tables. Line, bar, scatter, and pie charts are the most dominant chart types, covering more than 98.91\% of charts. The x-axes of bar and pie charts directly display the data values of their reference field one by one, which play the role of natural key. Y-axes display data size or trend, which plays the role of measure. And in some charts, the y-axis is plotted with multi-fields, which are measure pair.

2) Pivot dataset: There are 23,728 pivot tables (\eg, the one in \reffig{fig:overall}) created from 8,900 tables. Pivot table has ``rows'', ``columns'' and ``values''. Both the rows and columns 
%(correspond to any number of fields) 
hierarchically break down records into groups, which play the role of common breakdown. Its values 
%(correspond to one or more fields)
are for applying aggregation to each group, which plays the role of common measure and provides default aggregation functions.

\subsubsection{Web Tables}
Several web table (HTML table) datasets have been extracted in prior work, and the following datasets are used in this work:

(1) T2D dataset: T2D Gold Standard~\cite{ritze2017T2Dv2} is a dataset for evaluating matching systems on the task of matching Web tables to the DBpedia knowledge base. It contains schema-level correspondences between 1,724 Web tables (consisting of 7,705 fields) from the English-language subset of the Web Data Commons Web Tables Corpus~\cite{webtablecorpus} and DBpedia\footnote{https://www.dbpedia.org/}. In total, more than 2,000 fields are mapped to $\sim$290 DBPedia properties.
    
(2) TURL dataset: TURL~\cite{deng2020TURL} constructs a dataset based on the WikiTable corpus, and utilizes each cell hyperlink (link to Wikipedia pages) to get entity linking, column type, and relation extraction supervision labels. Our work utilizes TURL tables with column types that contain 2,368,990 columns from 403,450 tables.

\subsubsection{Synthetic Tables}
To utilize richer data and supervision labels, synthetic datasets are also taken into consideration.  SemTab challenge~\cite{jimenez2019semtab} aims at benchmarking systems dealing with the tabular data to knowledge graph matching problem, including the CEA task (matching a cell to a KG entity), the CTA task (assigning a semantic type to a column), and the CPA task (assigning a property to the relationship between two columns). This challenge provides large datasets automatically generated from the knowledge graph. SemTab 2020 dataset contains 131,289 tables and 68,001 tables with CPA task labels. To avoid data leakage, 17,995 schemata are extracted from 68,001 tables, and randomly keep one table in each schema. 

\subsection{Supervision Labels}
\label{sec:supervision}
Based on the above datasets, supervision labels are prepared in three ways: First, from the analysis artifacts created in downstream tasks including chart and pivot table; Second, from manual labels in the public dataset (\ie, T2D) and by ourselves; Third, from the information attached to the table (\ie, SemTab, TURL). \reftab{tab:corpus} shows supervision labels for each task from each dataset.

In \reftab{tab:corpus} rows 1-2, for the measure / dimension dichotomy in \refsec{sec:msr-dim}, the positive samples (measures) are the fields referenced by y-axis in charts and ``values'' in pivot tables, the negative samples (dimensions) are x-axis (except scatter and line charts) in charts and ``rows'' and ``columns'' in pivot tables.

For the common field roles in \refsec{sec:common-msr-dim}, the labels are slightly modified on measure / dimension dichotomy labels.  In \reftab{tab:corpus} row 4, for the common breakdown, positive samples come from the non-unique dimensions in pivot tables. In \reftab{tab:corpus} row 5, for natural key, samples come from dimensions in charts, and satisfy that 1) data values are unique 2) don't have multi-dimensions in the table. The negative samples are all other fields besides the positive samples in a given table. 

In \reftab{tab:corpus} row 8, for measure pair in \refsec{sec:msr-pair-type}, its positive labels come from pairs of measure fields referenced by the same chart axis. For each table, we randomly sample the same number of negative samples (numerical field pairs) as positive samples. 
In \reftab{tab:corpus} row 7, for measure type labels, we merge two label sources together: DBPedia property labels in T2D, and Wikidata property labels in SemTab.
We have organized the measure type taxonomy and map supervisions from the datasets, and details are shown in  \refsec{sec:app_msr_type}.
In row 6, for dimension type labels, we utilize column types in TURL\cite{deng2020TURL}. %628,254 columns from 397,098 tables for training, 13,025 (13,391) columns from 4,764 (4,844) tables for test (validation).

In \reftab{tab:corpus} row 9, for default aggregation ranking scores in \refsec{sec:default-agg}, we focus on the 9 most frequently used aggregation functions in pivot tables as shown in \reffig{fig:agg_funcs}. For a field, the actual aggregation applied by users has a score = 1, otherwise, unused functions are assigned with 0 score.

\subsection{Quality  Inspection}
To assess the quality of our corpus, we conducted a quality inspection with 5 experts who have analysis experience. We ensured accuracy by informing the annotators of the scoring standards during the annotation process. The score uses ordinal scale values(1, 0.75, 0.5, 0.25, 0) with corresponding definitions. Our results showed that the AnaMeta dataset labels got 0.97 (out of 1) on the macro average with manual annotation, which demonstrates that our corpus contains high-quality data and supervision. Details of inspection and results are shown in \refsec{sec:app-quality}.

% End within 4.5 pages

\section{KDF Framework}
\label{sec:method}
% \begin{figure*}[htp]
%     \centering
%     \includegraphics[width=2 \columnwidth]{figures/model1.png}
%     \caption{Metadata Model Architecture. \Circled{1} represents knowledge fusion module in \reffig{fig:knowledge_fusion}, and \Circled{2} represents distribution fusion module in \reffig{fig:distribution_fusion}.}
%     \label{fig:model_graph}
%     % \vspace{-2mm}
% \end{figure*}

% \begin{figure*}[tbp]
% 	\centering
% 	\subfloat[Knowledge Fusion Module. The module represents each sequence.]
% 	{\includegraphics[width=0.9\columnwidth]{figures/model2-1.pdf}\label{fig:knowledge_fusion}}\quad
% 	\subfloat[Distribution Fusion Module. The module represents each token.]
% 	{\includegraphics[width=0.9 \columnwidth]{figures/model2-2.pdf}\label{fig:distribution_fusion}}\\
% 	\caption{Metadata Model Architecture}
% 	\vspace{-2mm}
% \end{figure*}

As discussed in \refsec{sec:intro}, To improve the metadata understanding capability of tabular models, we propose KDF (Knowledge and Distribution Fusion) framework to infuse information. In this section, we provide a detailed description of Metadata model architecture as depicted in \reffig{fig:model_graph}.

\subsection{Overall Model Architecture}
\label{sec:pretrained}

Our model is based on the pre-trained tabular model and utilizes a transformer encoder on top. Because distribution and knowledge graph information describes both cells and columns, we design two encoders for KDF framework. The first encoder is sub-token or cell level (depending on the pre-trained tabular model), and the second one is column level. Details of KDF overall architecture are shown in \refsec{sec:app-overall}.

The pre-trained tabular model could be almost any transformer-based pre-trained models. In this work, we use TAPAS\cite{Herzig2020TAPASWS} and TABBIE\cite{Iida2021TABBIEPR} to illustrate effectiveness.

\subsection{Knowledge Fusion Module}
\label{sec:knowledge_fusion}
\begin{figure}[tbp]
    \centering
    \includegraphics[width=0.9\columnwidth]{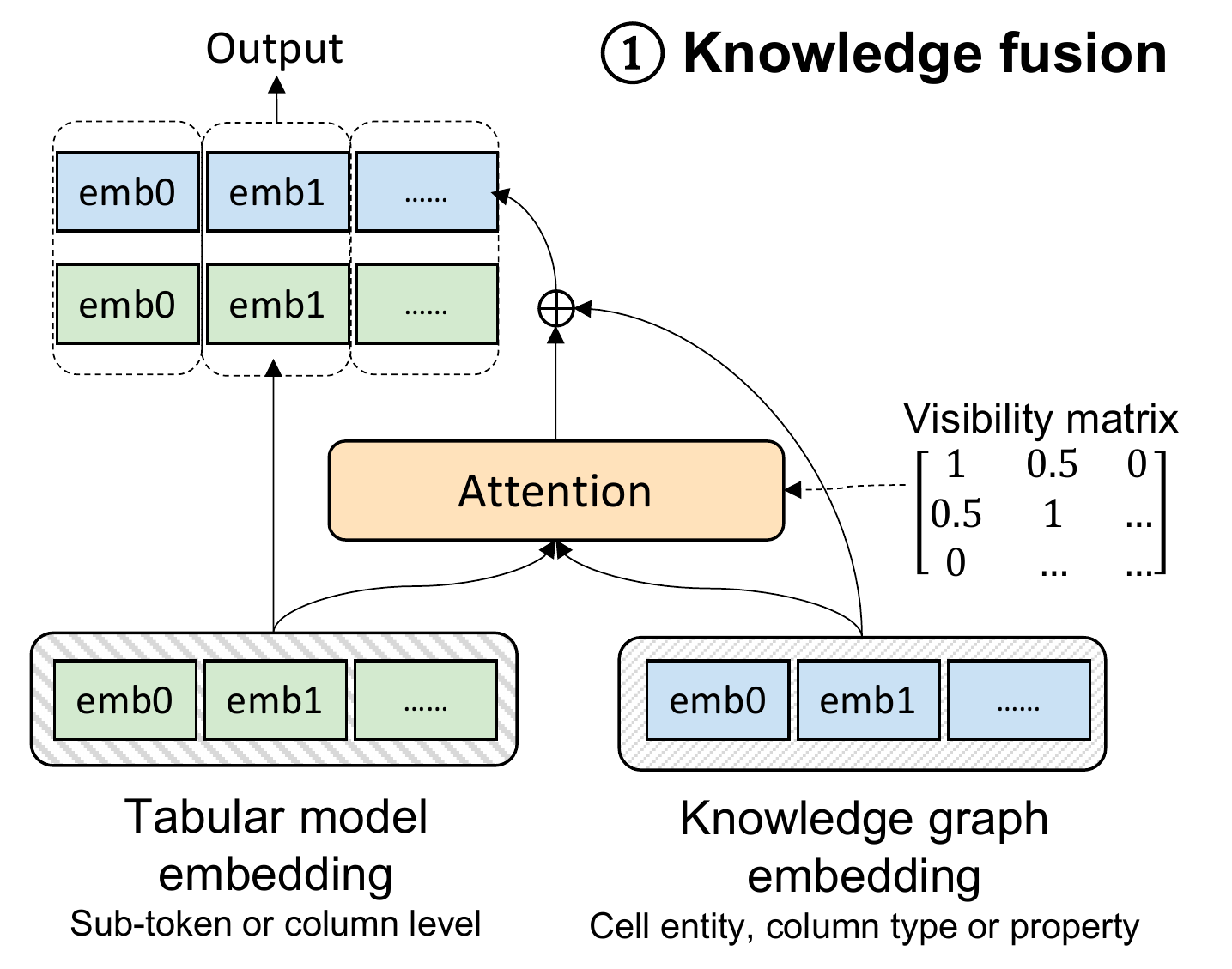}
	\caption{Knowledge Fusion Module in KDF Framework. The module represents each sequence.}
    \label{fig:knowledge_fusion}
	\vspace{-4mm}
\end{figure}
Knowledge graph information provides extra knowledge about tables and helps the model interpret tables. In this subsection, we describe how to get that information and how to use them.

Fusion module is illustrated in \reffig{fig:knowledge_fusion}. Model respectively fuses cell entity in cell level, column type, and property type in column level. Because one tabular token needs to pay attention to several entities, \eg, a cell token needs to pay attention to a specific entity in the same column or row, knowledge fusion module applies attention to fuse tabular embedding and knowledge graph embedding, as shown in the following equation.
\begin{equation*}
\small
\begin{split}
    &H  = W_3\ (\textup{Attention}(TOK, ENT, ENT) + ENT)\\
    &\textup{Attention}(Q, K, V) = W_2\ \textup{softmax}(QW_1K^T+\ln M)\ V
\end{split}
\end{equation*}
where, $TOK \in \mathbb{R}^{n\times d_{tok}}$ is the sequence of tabular model embedding, $ENT \in \mathbb{R}^{n\times d_{ent}}$ is the sequence of knowledge graph embedding, $W_1 \in \mathbb{R}^{d_{tok}\times d_{ent}}$, $W_2 \in \mathbb{R}^{d_{ent}\times d_{ent}}$ and $W_3 \in \mathbb{R}^{d_{ent}\times d_{h}}$ are trainable weights. $M \in \mathbb{R}^{n\times n}$ is visibility matrix.

The visibility matrix is to control whether an entity is visible to a tabular model token, as shown in \refequ{equ:visibility}. Besides, it only works in sub-token (or cell) level fusion.
% \vspace{-2mm}
\begin{equation}
\small
    M_{i,j}=\left\{
    \begin{array}{lll}
    1& \text{token $i$ and $j$ are from the same cell}\\
    m& \text{$i$ is only in the same column or row with $j$}\\
    0 & \text{others}
    \end{array}
    \right.
    \label{equ:visibility}
\end{equation}
where, $m \in (0, 1)$ is half visible hyper parameter.

After getting attentive entity embedding ($H$), we concatenate tabular model embedding ($TOK$) with it as the module's output.

\subsection{Distribution Fusion Module}
\label{sec:data_features}
It is hard for a pre-trained tabular model to capture the entire column distribution, which is important to analyze metadata tasks. To learn numerical distribution, we extract 31 data statistics features and 6 categories for each field~\cite{zhou2021charts}. 
In this module, tabular model embedding and field categories are added together after linear layer and embedding lookup respectively. The output of the module is the concatenation of the embedding and statistics features.
 Details of architecture and features are shown in \refsec{sec:app-df}.

\section{Downstream Interfaces}
\label{sec:interfaces}
Field metadata is formulated as classification problems, which can be effectively learned through the KDF framework. However, it remains a challenge to effectively incorporate this metadata into downstream applications. To address this challenge, we propose four interfaces including metadata IDs, embeddings, sentences, and  pre-training. These interfaces allow for the integration of metadata at different stages of the analysis process and facilitate its use in downstream applications.
\begin{table*}[!h]
  \centering
  \small
\centering
\caption{Benchmarks and KDF Framework Results on Metadata Tasks. All metric numbers are in \% and averaged over 3 runs. For each task the \textbf{bold} number is the best one. $\Delta$ means the best results of ``KDF'' metric minus the corresponding evaluation metric, and color formatting reacts value size. ``TML'' means traditional machine learning, ``TPLM'' means tabular pretrained language model, ``LLM'' means large language model.
}
\label{tab:eval_overall}
\label{tab:exp_result}%
\resizebox{\textwidth}{!}{
    \begin{tabular}{cl|cc|cccccc|cccccc|cc}
    \toprule
    \multicolumn{2}{c|}{\multirow{2}[4]{*}{Model}} & \multicolumn{2}{c|}{Msr } & \multicolumn{2}{c}{Natural Key} & \multicolumn{2}{c}{ Com. Breakdown} & \multicolumn{2}{c|}{ Com. Measure} & \multicolumn{2}{c}{Dim Type } & \multicolumn{2}{c}{Msr Type} & \multicolumn{2}{c|}{Msr Pair} & \multicolumn{2}{c}{Aggregation} \\
\cmidrule{3-18}    \multicolumn{2}{c|}{} & \multicolumn{1}{c}{Acc.} & \multicolumn{1}{c|}{$\Delta$} & \multicolumn{1}{c}{HR\@1} & \multicolumn{1}{c}{$\Delta$} & \multicolumn{1}{c}{HR@1} & \multicolumn{1}{c}{$\Delta$} & \multicolumn{1}{c}{HR@1} & \multicolumn{1}{c|}{$\Delta$} & \multicolumn{1}{c}{Acc.} & \multicolumn{1}{c}{$\Delta$} & \multicolumn{1}{c}{Acc.} & \multicolumn{1}{c|}{$\Delta$} & \multicolumn{1}{c}{Acc.} & \multicolumn{1}{c|}{$\Delta$} & \multicolumn{1}{c}{Acc.} & \multicolumn{1}{c}{$\Delta$} \\
    \midrule
    \multicolumn{2}{c|}{Rule based} & 96.68  & \cellcolor[rgb]{ .996,  .933,  .933}-2.26  & 88.42  & \cellcolor[rgb]{ .988,  .804,  .804}-6.62  & 49.72  & \cellcolor[rgb]{ .976,  .533,  .541}-15.77  & 67.02  & \cellcolor[rgb]{ .992,  .859,  .863}-4.71  & 12.99  & \cellcolor[rgb]{ .973,  .412,  .42}-84.98  & 25.91 & \cellcolor[rgb]{ .973,  .412,  .42}-53.41  & 32.04  & \cellcolor[rgb]{ .973,  .412,  .42}-46.97  & 90.53  & \cellcolor[rgb]{ .996,  .969,  .969}-0.95  \\
    \midrule
    \multicolumn{1}{c}{\multirow{2}[2]{*}{TML}} & GBDT  & 96.94  & \cellcolor[rgb]{ .996,  .941,  .941}-1.99  & 89.38  & \cellcolor[rgb]{ .992,  .831,  .835}-5.66  & 55.50  & \cellcolor[rgb]{ .984,  .706,  .71}-9.99  & 66.84  & \cellcolor[rgb]{ .992,  .855,  .855}-4.90  & 24.44  & \cellcolor[rgb]{ .973,  .412,  .42}-73.53  & 63.86  & \cellcolor[rgb]{ .976,  .545,  .549}-15.47  & 62.55  & \cellcolor[rgb]{ .976,  .514,  .522}-16.46  & 90.55  & \cellcolor[rgb]{ .996,  .969,  .973}-0.94  \\
          & RF    & 98.07  & \cellcolor[rgb]{ .996,  .973,  .973}-0.87  & 92.75  & \cellcolor[rgb]{ .996,  .929,  .933}-2.29  & 59.47  & \cellcolor[rgb]{ .988,  .82,  .824}-6.01  & 67.13  & \cellcolor[rgb]{ .992,  .863,  .863}-4.61  & 45.56  & \cellcolor[rgb]{ .973,  .412,  .42}-52.41  & 66.32  & \cellcolor[rgb]{ .98,  .616,  .62}-13.00  & 63.76  & \cellcolor[rgb]{ .976,  .549,  .557}-15.25  & 90.84  & \cellcolor[rgb]{ .996,  .98,  .98}-0.65  \\
    \midrule
    \multicolumn{1}{c}{\multirow{3}[2]{*}{TPLM}} & TURL  & 97.65  & \cellcolor[rgb]{ .996,  .961,  .961}-1.28  & 92.65  & \cellcolor[rgb]{ .996,  .929,  .929}-2.39  & 62.38  & \cellcolor[rgb]{ .992,  .906,  .906}-3.11  & 70.37  & \cellcolor[rgb]{ .996,  .957,  .957}-1.37  & 96.55  & \cellcolor[rgb]{ .996,  .957,  .957}-1.42  & 66.92  & \cellcolor[rgb]{ .98,  .631,  .639}-12.41  & 69.38  & \cellcolor[rgb]{ .984,  .714,  .718}-9.63  & 91.15  & \cellcolor[rgb]{ .996,  .988,  .988}-0.33  \\
          & TAPAS & 97.09  & \cellcolor[rgb]{ .996,  .945,  .945}-1.84  & 93.79  & \cellcolor[rgb]{ .996,  .961,  .961}-1.25  & 58.62  & \cellcolor[rgb]{ .988,  .796,  .8}-6.87  & 70.01  & \cellcolor[rgb]{ .996,  .949,  .949}-1.73  & 96.21  & \cellcolor[rgb]{ .996,  .945,  .945}-1.76  & 78.49  & \cellcolor[rgb]{ .996,  .973,  .973}-0.83  & 75.20  & \cellcolor[rgb]{ .992,  .886,  .886}-3.81  & 89.96  & \cellcolor[rgb]{ .996,  .953,  .953}-1.52  \\
          & TABBIE & 97.49  & \cellcolor[rgb]{ .996,  .957,  .957}-1.45  & 94.42  & \cellcolor[rgb]{ .996,  .98,  .98}-0.62  & 62.36  & \cellcolor[rgb]{ .992,  .906,  .906}-3.13  & 71.16  & \cellcolor[rgb]{ .996,  .98,  .98}-0.58  & 95.34  & \cellcolor[rgb]{ .996,  .922,  .922}-2.63  & 77.39  & \cellcolor[rgb]{ .996,  .941,  .941}-1.94  & 77.70  & \cellcolor[rgb]{ .996,  .961,  .961}-1.32  & 88.43  & \cellcolor[rgb]{ .992,  .91,  .91}-3.05  \\
    \midrule
    \multicolumn{1}{c}{\multirow{2}[2]{*}{LLM}} 
          & GPT-3.5 0-shot & 87.18  & \cellcolor[rgb]{ .98,  .655,  .659}-11.75  & 88.39  & \cellcolor[rgb]{ .988,  .808,  .812}-6.65  & 41.51  & \cellcolor[rgb]{ .98,  .58,  .588}-23.98  & 50.18  & \cellcolor[rgb]{ .973,  .412,  .42}-21.56  & 46.11  & \cellcolor[rgb]{ .973,  .412,  .42}-51.86  & 35.10  & \cellcolor[rgb]{ .973,  .412,  .42}-44.22  & 51.45  & \cellcolor[rgb]{ .973,  .412,  .42}-27.57  & 77.46  & \cellcolor[rgb]{ .98,  .584,  .588}-14.02  \\
          & GPT-3.5 1-shot & 91.14  & \cellcolor[rgb]{ .988,  .765,  .769}-7.79  & 82.47  & \cellcolor[rgb]{ .98,  .639,  .643}-12.57  & 44.80  & \cellcolor[rgb]{ .988,  .796,  .8}-20.69  & 57.30  & \cellcolor[rgb]{ .973,  .412,  .42}-14.43  & 46.38  & \cellcolor[rgb]{ .973,  .412,  .42}-51.59  & 29.32 & \cellcolor[rgb]{ .973,  .412,  .42}-50.00  & 55.36  & \cellcolor[rgb]{ .973,  .412,  .42}-23.66  & 75.66  & \cellcolor[rgb]{ .973,  .475,  .482}-15.82  \\
    \midrule
    \multicolumn{1}{c}{\multirow{2}[2]{*}{KDF}} & KDF+TAPAS & 98.45  & \cellcolor[rgb]{ .996,  .984,  .984}-0.48  & \textbf{95.04} & 0.00  & 64.03  & \cellcolor[rgb]{ .996,  .957,  .957}-1.46  & 71.09  & \cellcolor[rgb]{ .996,  .98,  .98}-0.65  & \textbf{97.97} & 0.00  & \textbf{79.32} & 0.00  & 77.50  & \cellcolor[rgb]{ .996,  .953,  .953}-1.52  & \textbf{91.48} & 0.00  \\
          & KDF+TABBIE & \textbf{98.93} & 0.00  & 94.99  & \cellcolor[rgb]{ .996,  .996,  .996}-0.06  & \textbf{65.49} & 0.00  & \textbf{71.74} & 0.00  & 97.33  & \cellcolor[rgb]{ .996,  .98,  .98}-0.64  & 78.74  & \cellcolor[rgb]{ .996,  .98,  .98}-0.58  & \textbf{79.01} & 0.00  & 90.58  & \cellcolor[rgb]{ .996,  .973,  .973}-0.90  \\
    \bottomrule
    \end{tabular}%
    }
  \label{tab:addlabel}%
  % \vspace{-4mm}
\end{table*}%

\textbf{Metadata IDs Interface} provides downstream applications with the classification results of field metadata tasks. This interface can be used to provide more specific and targeted uses of field metadata. For example, IDs can be used as rules to limit the scope of downstream searches.  During QuickInsights\cite{ding2019quickinsights}, different data mining strategies are applied based on whether a field is classified as a measure or dimension.  Additionally, ids can be used as tags in language models. 

\textbf{Metadata Embeddings Interface} provides downstream applications with each column embeddings, which are augmented with metadata knowledge. These embeddings can represent columns in a continuous, dense, and low-dimensional vector space. Thus they can be used in a variety of data analysis tasks to classify columns, generate analysis according to the column, and so on. Additionally, they can be used as features in downstream machine learning models.

\textbf{Metadata Sentences Interface} provides downstream applications with string sentences, which describe the metadata knowledge associated with a field. These sentences can provide a human-readable summary of the metadata. As language models with a seq2seq structure have become popular in recent years, using sentences as an interface to combine different models and make use of their knowledge is a current trend. Therefore, by using metadata sentences, it is possible to easily incorporate metadata into various downstream tasks.

\textbf{Metadata Pretraining Interface} provides metadata tasks as pre-training or continue pre-training objectives. This allows for the direct incorporation of metadata knowledge into downstream models during the training stage. Additionally, it should be noted that metadata can be formulated as question-answer tasks to make it more adaptable to a wider range of language models, rather than just classification tasks.

\section{Experiments}
\label{sec:exp}

We conducted three parts of experiments to evaluate field metadata tasks on the AnaMeta dataset and demonstrate how to incorporate analysis knowledge. First, a variety of models are evaluated on all metadata tasks as a benchmark. Second, the KDF framework is evaluated on all metadata tasks 
% to demonstrate the usefulness of distribution and knowledge fusion. 
Finally, we evaluate the performance of downstream applications using four interfaces. All the experiments are run on Linux machines with 448 GB memory, and 4 NVIDIA Tesla V100 16G-memory GPUs.

\vspace{-1mm}
\subsection{Benchmark}
\label{sec:benchmark}
To compare the performance of existing models on all tasks, we use four kinds of baselines -- rule-based baseline, traditional machine learning baselines (GBDT and Random Forest), pre-trained tabular model baselines (TURL), and large language model (GPT-3.5: text-davinci-003). Because there is no direct existing experiment on metadata tasks, we slightly change the above models to adapt to our tasks. Details of implementations and GPT-3.5 prompt engineering are shown in \refsec{sec:app-baseline}.
As described in \reftab{tab:eval_overall}, we find the following insights:
% Rule-based

\textit{Semantic information captured by the pre-trained tabular models brings great gain to metadata tasks.} Tabular pre-trained language model outperforms rule-based and traditional machine learning approaches, especially on common field roles and semantic field types (\eg dimension type outperforms about 80\% and 60\%). Traditional machine learning is good at representing field distribution, while it lacks semantic information and entire table understanding. Pre-trained tabular model fills the gap. 
% \Eg, TABBIE achieves the top one on all common field roles tasks among benchmarks and outperforms traditional machine learning by 6.86\% and 2.89\% on common breakdown. 
% It outperforms the rule-based baseline by more than 10\% on each task. Because common tasks can not be directly identified by field type or positional information, they are not easy for rules.

In our preliminary experiments, we investigated the capability of large language models (LLMs) to extract metadata from tables. 
As shown in \reftab{tab:exp_result}, our results indicate that GPT-3.5 has the ability to extract metadata, though it is not yet ideal. 
One limitation is that GPT-3.5 may not have been exposed to enough metadata during its training, which can result in performance that is worse than traditional heuristic rules.
However, we found that providing GPT-3.5 with one example improved its performance on more than half of the metadata tasks, suggesting that with proper in-context learning, GPT-3.5 could have even better metadata extraction capabilities.
Additionally, our initial exploration with naive prompt designs may not have fully leveraged the knowledge present within the large language models.
Further research in these areas is warranted.

% \subsection{Performance Comparisons}
% \label{sec:performance}
\vspace{-1mm}
\subsection{KDE Framework}
\label{sec:main_results}
Experiments are conducted on all metadata tasks. We respectively use TAPAS~\cite{Herzig2020TAPASWS} and TABBIE~\cite{Iida2021TABBIEPR} as a pre-trained tabular model to illustrate the effectiveness of KDF framework. Details of KDF experiment setting are shown in \refsec{sec:app-KDF-exp}. As described in \reftab{tab:eval_overall}, we find the following insights across tasks.

\textit{Successful distribution and knowledge fusion brings a better representation of fields.} KDF models exceed the performance of the other pre-trained tabular models, especially on common field roles, measure type, and measure pair tasks. 
% On common breakdown task, KDF(TABBIE) outperforms TABBIE by 3.13\%. 
On the measure type task, KDF (TAPAS) outperforms TURL by 12.41\%. Our KDF framework explicitly fuses distribution and knowledge information. Thus, it can better represent the whole field and integrate useful external information. More analysis of the two fusions is described in ablation experiments (\refsec{sec:ablation}).

On the dimension type task, KDF (TAPAS) achieves the top result and improves over the performance of  TURL by 1.42\%. It's worth noting that this task is one of TURL's original tasks, while our Metadata model still has advantages. It benefits from better knowledge fusion and pre-training model representation.

\subsection{Downstream Interfaces}
\label{sec:exp-downstreamtask}

\begin{table}[htbp]
\vspace{-2mm}
\small
\centering
\caption{Metadata IDs and Embeddings 
Interfaces Results on Table2Chart. All metric numbers are averaged over 3 runs.}
\begin{tabular}{c|ccc}
\toprule
Interface          & Precision & Recall   & F1       \\ \midrule 
w/o Metadata        & 80.60\%  & 77.28\% & 78.91\% \\
Metadata IDs & 81.45\%  & \textbf{77.73\%} & 79.55\% \\ 
Metadata Embeddings & \textbf{82.44\%}  & 77.56\% & \textbf{79.92\%} \\
\bottomrule
\end{tabular}
\label{tab:table2charts}
\end{table}

% Table generated by Excel2LaTeX from sheet 'Sheet1'
\begin{table}[htbp]
\small
  \centering
  \caption{Metadata Sentences and Pre-training Interfaces Results on TableQA. All metric numbers are averaged over 3 runs.}
    \begin{tabular}{c|cc}
    \toprule
       Interface   & \multicolumn{1}{l}{HybridQA} & \multicolumn{1}{l}{WikiTQ} \\
    \midrule 
    w/o Metadata & 53.69\% & 36.70\% \\
    Metadata Sentences & \textbf{54.59\%} & \textbf{37.83\%} \\
    Metadata Pre-training & 53.62\% & 36.14\% \\
    \bottomrule
    \end{tabular}%
    % \vspace{-4mm}
  \label{tab:tableqa}
\end{table}%

In order to evaluate the performance of four interfaces introduced in \refsec{sec:interfaces}, and demonstrate the importance of field metadata, we conducted experiments using interfaces in several downstream analysis tasks.
For the Metadata IDs and embeddings interfaces, which take column features (embeddings) as input, we chose visualization generation tasks and applied the interfaces to the Table2Charts model~\cite{zhou2021charts}. For the metadata sentences and pre-training interfaces, which are suitable for tasks solved by pre-trained language models with string sentences as inputs, we chose the popular TableQA task and applied the interfaces to the UnifiedSKG framework~\cite{UnifiedSKG}. Results can be found in \reftab{tab:table2charts} and \reftab{tab:tableqa}. More implementation details can be found in \refsec{sec:downstream}.

\textit{Field metadata knowledge can improve downstream analysis tasks when used with the appropriate interface.} As seen in \reftab{tab:table2charts}, the Metadata embeddings interface outperforms the baseline with a 1.84\% increase in precision. Additionally, in Table~\ref{tab:tableqa}, the metadata sentences interface improves the TableQA task, even though this task does not directly use metadata as output. These results highlight the importance and necessity of field metadata knowledge learned from field metadata tasks.

\textit{Different downstream tasks may benefit from different interfaces depending on factors, such as input, task characteristics, model characteristics, \etc} In addition to downstream task inputs that strictly limit the choice of interfaces, the characteristics of the task and model are also important. In \reftab{tab:tableqa}, Metadata sentences interface can boost the task while metadata pre-training gives the model a bad influence. Metadata tasks have different logic of reasoning with TableQA tasks, although it can help understand tables. When using metadata task as pre-training objectives, it destroys the original logic of reasoning and brings side effects.

% End within 7.5 pages

\section{Related Work}
\vspace{-1mm}
\subsection{Table Interpretation}
\label{sec:table_interpretation}
\label{sec:related_sem_tab}
There is a long line of work trying to understand tables symbolically, especially for entity or content tables where we can conduct entity linking, column type annotation, and relation extraction~\cite{Wang2012UnderstandingTO,Kacprzak2018MakingSO,hulsebos2019sherlock,cutrona2020semtab,wang2021TCN}. Some domain ontology or knowledge graph, such as DBPedia~\cite{lehmann2015dbpedia} and Wikidata~\cite{vrandevcic2014wikidata}, is often provided for alignment. 
Column type annotation and relation extraction are related to our measure / dimension (\refsec{sec:msr-dim}) and field type (\refsec{sec:msr-pair-type}) classification. However, most previous work focus on the entity type of the columns~\cite{hulsebos2019sherlock,deng2020TURL}, and few public datasets provide real-world tables with rich labels of measurements~\cite{ritze2017T2Dv2,cutrona2020semtab}. 

\vspace{-1mm}
\subsection{Pre-trained Tabular Models}
Pre-trained language models~\cite{devlin2019bert, brown2020language} are widely used in NLP tasks. Recently also emerge several pre-trained tabular models with transformer as the primary backbone~\cite{table_pre-training}, such as, TAPAS~\cite{Herzig2020TAPASWS}, TABBIE~\cite{Iida2021TABBIEPR} and TURL~\cite{deng2020TURL}.
The semantic information within those models contains an amount of metadata knowledge, however, the models still lack the ability to understand things like table distributions, which we demonstrate in \refsec{sec:exp}

% TaBERT~\cite{yin20TaBERT} and TAPAS~\cite{Herzig2020TAPASWS} serialize the table and jointly pre-train the text-table pairs with MLM objective to resolve the table QA problem. However, their representation capability is limited by the expensive computation due to table serialization.
% TABBIE~\cite{Iida2021TABBIEPR} remedies the issue by leveraging two transformers to encode rows and columns independently to reduce the length of inputs.
% TURL~\cite{deng2020TURL} proposes a structure-aware Transformer with  Masked Entity Recovery (MER) objective and injects entity knowledge in relational Web Tables~\cite{deng2020TURL}. 

\section{Conclusion}
\vspace{-1mm}
% In this paper, we propose the novel analysis metadata for tabular data analysis and collected a large corpus with supervision by using smart supervisions from downstream tasks, public datasets, and our manual labels. Then we propose Metadata model to understand field distribution and utilize knowledge graph information. We conduct several experiments to illustrate the importance of metadata tasks and the effectiveness of Metadata model. Analysis metadata and our experiences with tabular model design could benefit a wide variety of downstream tasks.

% Future work are as following: (1) There are far more types of analysis metadata to be discovered and inferred. 
% %Inspired by data profiling metadata, dependency between multi-fields in one table plays an important role. There are several common dependencies or relationship among columns. How to identify them is a future work.
% (2) Understanding numbers in table and design proper pre-train objectives is still a challenge for us to explore. Further explorations on semantic and distributional features are still needed.
% (3) Utilizing existing knowledge graph information to interpret and analysis table is important and promising. We make the first attempt in this work, and it's still a future work.

In conclusion, this paper has presented the AnaMeta dataset, a collection of tables with derived supervision labels for four types of commonly used field metadata. We have also proposed a multi-encoder framework, called KDF, which 
% improves the metadata understanding capability of tabular models by 
incorporates distribution and knowledge information. Additionally, we have proposed four interfaces for incorporating field metadata into downstream analysis tasks. Through evaluations of a wide range of models, we have shown the importance of accurate metadata understanding for tabular data analysis and the effectiveness of the KDF framework and interfaces in improving the performance of downstream tasks.

\section*{Limitations}
The type of field metadata tasks is limited in this paper and it can be explored more.
There are far more types of analysis metadata to be discovered and inferred. On the one hand, inspired by data profiling metadata, the dependency between multi-fields in one table plays an important role. There are several common dependencies or relationships among columns. How to identify them is future work.
On the other hand, in \reftab{tab:msr-taxonomy} only a limited taxonomy is provided. A more comprehensive one is future work.

% Tabular Model Improvements.
% As discussed in \refsec{sec:exp}, several existing approaches on tabular modeling all have improvement space in the future. There is still an efficiency gap for real-world deployment of large pre-trained tabular models in products. 
% Understanding numbers in tables and designing proper pre-train objectives is still a challenge for us to explore. 
% Meanwhile, feature engineering seems inevitable in tabular models at current stage. 
% Further explorations on semantic and distributional features are still needed.

% Utilizing existing knowledge graph information to interpret and analyze table is important and promising. We make the first attempt in this work, and it's still a future work.

Our initial research explored the ability of large language models (LLMs) to extract metadata from tables. The results were not optimal, likely due to a lack of exposure to metadata during the training process of the LLM and limitations in the design of the prompts used. Further investigation is necessary to improve the performance of LLMs in extracting metadata from tables.

\section*{Ethical Statements}
We collect AnaMeta dataset from 3 source datasets. For spreadsheet dataset, we crawl them from websites, which means original spreadsheet are public. Thus, We believe there is no privacy issue related to this dataset. For web tables and synthetic tables, we apply public datasets and follow their License. 

% Entries for the entire Anthology, followed by custom entries

\bibliography{anthology,references}
\bibliographystyle{acl_natbib}
\clearpage
\appendix

\label{sec:appendix}
\section{Related Work}
\label{sec:app-related}
\subsection{Dimensional Modeling and Metrology}
The terms ``measure'' and ``dimension'' have their roots in dimensional modeling from data warehousing and business intelligence~\cite{golfarelli1998dimensional}. In relational and multidimensional databases, dimensional models are implemented as star schemas %(where a fact table containing measures is joined with dimension tables by primary keys) 
and online analytical processing (OLAP) cubes~\cite{kimball2013data}.
Our definition of ``dimension'' extends the concept in dimensional modeling. As we will discuss in \refsec{sec:msr-dim}, it contains primary keys and natural keys in addition to the dimension attributes. %our dimension not only includes the dimension attributes, % (which serve as the primary source of query constrains, groupings, and report labels), 
%but also contains primary keys and natural keys. % This is a common practice in real-world data analysis products such as Excel and Tableau.
Most of our analysis metadata involves measures. As the scientific study of measurements, Metrology includes the definition of quantities and units of measurement. In \refsec{sec:msr-pair-type}, we will define our measure types with common units from the International System of Units (SI)~\cite{bidpe2019si,iso2021SI}, which is a widely accepted metric system.% The International System of Units (SI)~\cite{bidpe2019si,iso2021SI} is awidely accepted metric system. In \refsec{sec:msr-pair-type}, we will define our measure types with common units from SI.

% Data warehouse, as an integral system for data storing and analysis, has been devised to deliver crucial and easily-accessible data report for business intelligence and other fields. 
% As defined in data warehouse, measures and dimensions are
% significant concepts to distinguish the inherent characteristic of the attributes

% logical design to be more
% specifically aimed at the efficiency required by data warehousing applications continuously valued (typically numerical) attributes which describe the fact from different
% points of view; for instance, each sale is measured by its revenue. Dimensions are discrete
% attributes which determine the minimum granularity adopted to represent facts; typical
% dimensions for the sale fact are product, store and date

\subsection{Data Profiling}
\label{sec:data_profiling}
~\cite{abedjan2018dataprofiling}  has the similar term ``metadata''. These metadata, such as statistics about the data or dependencies among columns, can help understand and manage new datasets. %Exsiting metadata activities including single metadata profiling and dependencies. 
% Analysis metadata is proposed for further analysis tasks based on part of data profiling metadata. The details about their relationship for each task are discussed in \refsec{sec:app-related}

Analysis metadata is proposed for further analysis tasks based on part of data profiling metadata. The details about their relationship for each task are as follows:
% \begin{itemize}
(1) Unique column combination and primary key. There is plenty of work to explore them, while key with semantics is not their focus. \cite{bornemann2020natural} proposed natural key based on primary key, and they use engineered features and Random Forest to solve the problem. 
    
(2) Identifying semantic domain of a column. \cite{zhang2011automatic} first propose semantic domain labeling by clustering columns. %that have the same meaning across the tables of a database. And \cite{zhang2013infogather+} extend to identifying same semantic domain columns with different units. 
\cite{vogel2011instance} matches columns to pre-defined semantics 
%from the person domain 
with specific features and Naive Bayes. This track of works evolves towards column type identification in table interpretation, which we discuss in \refsec{sec:table_interpretation}. 
    
(3) Quantity name recognition: This activates detecting the quantity name for a column, which is highly correlated with measure type in analysis metadata. \cite{sarawagi2014open} point out that unit extraction is a significant step for queries on web tables, and design unit extractors for units in column names by developing a unit catalog tree. \cite{yi2018recognizing} extends to inferring unknown units with extracted feature and Random Forest. However, those existing works heavily depend on units appearing in the table, so we propose measure types that can also identify measure fields without units and property.
% \end{itemize}

State-Of-Art of those related works often extracts specific features and uses traditional machine learning to solve the problem, which lacks further semantic representation with the pre-training model.

\subsection{Downstream Analysis Tasks}
\label{sec:downstream-tasks}
Lots of intelligent data analysis features could benefit from analysis metadata. Typical examples include automatic insights discovery~\cite{ding2019quickinsights,law2020vis}, chart and pivot table recommendations~\cite{Zhou2020Table2AnalysisMA,zhou2021charts,aoyu2021survey}, Text2SQL and query recommendations~\cite{dong2016nl2sql,km2021survey,yu2021grappa}, table expansion~\cite{Zhang2017EntiTablesSA,Zhang2019AutocompletionFD}, \etc Most of these tasks involve searching, enumerating, and comparing in a large space. Analysis metadata could help 
narrow down possible candidates (prioritized searching order) and provide good ranking references.

\section{Problem Definition}
\subsection{Measure / Dimension Dichotomy}
First, not all numerical fields are measure fields. In \reftab{tab:msr_case}, ``Style'' fields consist of categorical numbers, thus is dimension.  ``ID'' and ``QP Code'' fields represent keys, thus are dimensions. %An ML model should understand both linguistic semantics and number distribution to distinguish such fields. 
Second, there exists a weak dependency between field positions and roles. Starting from the left, usually, key dimensions come before breaking down dimensions, and measures come after dimensions.
%Sometimes in real-world dimensions are arranged in succession towards the left side of a table, while measures appear one-after-another towards the right end. 
An ML model should take vague hints among fields into account.

\begin{table}%[htbp]
  \centering
  \caption{Numerical Dimension. }
  \resizebox{.6\linewidth}{!}{%
    \begin{tabular}{l|l}
    \toprule
    Header  & Records \\
    \midrule
    Class  & 1,2,3,4,5... \\
    Rank  & 11,12,13,14,15... \\
    ID    & 9131115,22112723,1111145,30320912... \\
    QP Code & 1256,1245,1237,2134... \\
    Style & 4,4,2,2,2... \\
    %CodeAllocatedType & 401,401,101,101,101... \\

    \bottomrule
    \end{tabular}%
  }
  \label{tab:msr_case}%
  \vspace{-5mm}
\end{table}%

%  Measure / dimension dichotomy cannot simply be equated to a numerical / string dichotomy. Rule-based baselines directly classify the field by numerical and string. And as shown in \reftab{tab:exp_result}, Metadata (TABBIE) model reaches 98.48\%,  outperforms rule-based model by 3.83\%, and exceeds traditional machine learning model and TURL. 

% Measure / dimension dichotomy is the cornerstone of all tasks, and its accuracy affects all the tasks. Thus, we focus on its case study.
% The key point of this task is how to identify the dimension field from numerical fields. \reftab{tab:msr_case} shows some cases that Metadata successfully identifies dimension from numerical fields. ``ID'' and ``QP Code'' show that when numerical values are hard to identify the field
% , Metadata learns semantic representation from the pre-training model and successfully identifies the field. ``Style'' shows that when the semantic of the header is hard to identify, Metadata learns the numerical distribution to identify the group-by field.

\subsection{Common Field Roles}
It's worth noting that there are several existing works on ``primary key'', while ``natural key'' is different from ``primary key'' as described in \refsec{sec:data_profiling}. Both of them are chosen to represent records, while ``primary key'' focuses on unique (\eg, ID) and ``natural key'' focuses on semantic terms (\eg, name). 
%We adopt the name ``natural key'' here because in majority of the cases only the top key dimension is used in analysis. It echos the primary key and natural key concepts in databases~\cite{kimball2013data}. 
Sometimes it is also called key / core / subject / name / entity column in relational tables~\cite{ritze2017T2Dv2,zhang2020webtable}. As discussed in \refsec{sec:related_sem_tab}, natural key can be used for entity linking and other table understanding tasks.%column type identification、relation extraction and row population
% \end{definition}

\subsection{Dimension Type}
\label{sec:app-dimtype}
We adopt TURL~\cite{deng2020TURL} column types as dimension type. It contains 255 types and the most frequency dimension types are shown in \reftab{tab:dim_type}.
\begin{table}[htbp]
 \small
  \centering
  \caption{The Most Frequency Dimension Types.}
  \resizebox{.9\linewidth}{!}{%
    \begin{tabular}{ll}
    \toprule
    \multicolumn{2}{c}{Dimension type} \\
    \midrule
    people.person & government.political\_party \\
    location.location & location.administrative\_division \\
    organization.organization & sports.sports\_league\_season \\
    sports.sports\_team & soccer.football\_player \\
    sports.pro\_athlete & sports.sports\_league \\
    soccer.football\_team & government.politician \\
    time.event & film.film \\
    location.country & business.business\_operation \\
    location.citytown & ... \\
    \bottomrule
    \end{tabular}%
    }
  \label{tab:dim_type}%
%   \vspace{-3mm}
\end{table}%

\subsection{Measure Type}
\label{sec:app_msr_type}

Each type represents a magnitude, and each type is mutually exclusive. Each measure type corresponds to a set of convertible units (see ``Common Units'' in \reftab{tab:msr-taxonomy}), and highly correlated concepts (see ``Common Examples'' in \reftab{tab:msr-taxonomy}). ``Dimensionless'' category with 6 types is summarized in the taxonomy. Different from the existing magnitude or units taxonomy, there is plenty of measure without units and corresponding measure types are important in analysis. Thus we summarize the most common mutually exclusive 6 dimensionless measure types.

\begin{table*}
\centering
\caption{Measure Type Taxonomy and Details.}
\small
\label{tab:msr-taxonomy}
\begin{tabular}{llll} 
\toprule
Category                       & Type                 & Common Examples                          & Common Units             \\ 
\midrule
\multirow{6}{*}{Dimensionless} & Count (Amount)       & Population, daily passenger flow, ...                               &                                   \\ 
\cline{2-4}
                               & Ratio                & Percentage, change rate, proportion, ...      &                                  \\ 
 \cline{2-4}
                               & Angle                & Angle, longitude, latitude ...      &                                 \\ 
\cline{2-4}
                               & Factor / Coefficient & Coefficient of thermal expansion, drag coefficient, ...                                         &                                \\ 
\cline{2-4}
                               & Score                & Rating, exam score, indicator (index), ...    &                              \\ 
\cline{2-4}
                               & Rank                 & University ranking, projected GDP ranking, ...                                         &                                 \\ 
\hline
Money                          &                      & Sales, asset, income, revenue, cost, GDP, ... & \$, €, £, ...            \\ 
\hline
Data/file size                 &                      & Memory size, disk size, ...                   & GB, kb, ...                  \\ 
\hline
\multirow{2}{*}{Time}          & Duration             & Age, runtime, time length, ...                & s, min, hr, d, yr, ...   \\ 
\cline{2-4}
                               & Frequency            & Audio frequency, rotational speed, ...                                         & Hz, RPM, ...                          \\ 
\hline
\multirow{9}{*}{Scientific}    & Length               & Length, width, elevation, depth, height, ...  & m, cm, yard, feet, ...           \\ 
\cline{2-4}
                               & Area                 & Surface area, gross floor area, ...                                         & m$^2$, acre, ...         \\ 
\cline{2-4}
                               & Volume (Capacity)    & Vital capacity, water capacity, ...                                         & m$^3$, L, ...         \\ 
\cline{2-4}
                               & Mass (Weight)        & Body weight, salt consumption, ...                                         & kg, lbs, ...      \\ 
\cline{2-4}
                               & Power                & Source power, rated power, ...                                          & kW, ...   \\ 
\cline{2-4}
                               & Energy               & Calories, energy consumption, ...                                         & J, kcal, ...  \\ 
\cline{2-4}
                               & Pressure             & Atmospheric pressure, blood pressure, ...                                         & Pa, mmHg, ...   \\ 
\cline{2-4}
                               & Speed
                               & Velocity, average speed, ...
                               & m/s, km/h, ...         \\
\cline{2-4}
                               & Temperature             & Effective temperature, melting point, boiling point, ... 
                               & $^\circ C$, K, ...\\
\bottomrule
\end{tabular}
\end{table*}

\begin{figure}[htbp]
    \centering
    \includegraphics[width=1\columnwidth]{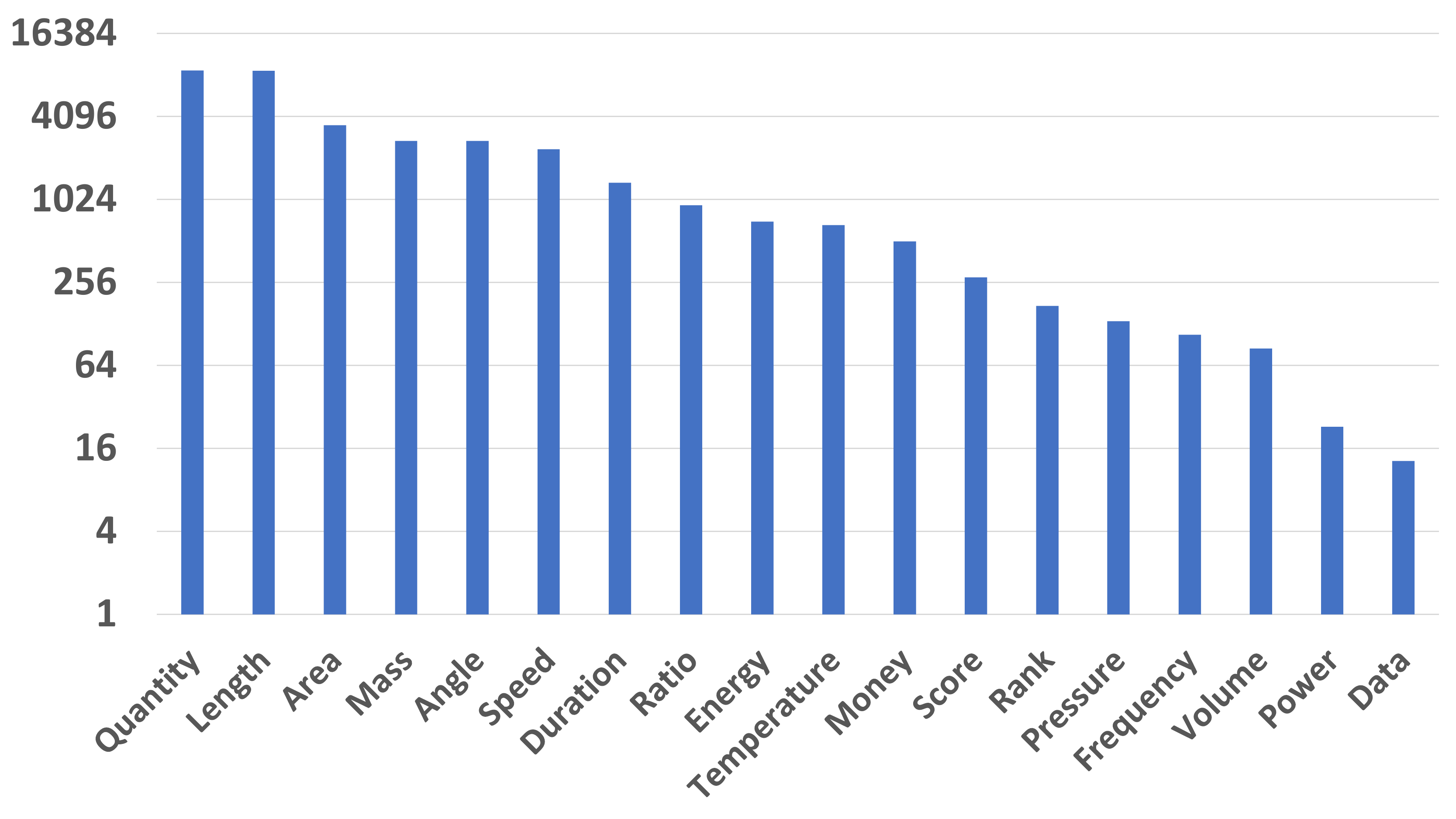}
    \caption{Measure Type. (Taking the logarithm of y-axis with base 4)}
    \label{fig:msr_type}
    \vspace{-3mm}
\end{figure}

The field still has a measure type, when values of the measure field are not entity or property in KG. Measure type is important to all common measure fields. But the existing column type identification and relation extraction task need table values to be entities or properties in the knowledge graph. 
%In other words, in addition to Wikipedia tables, measure type also is identified in the other relational tables which don't match with knowledge graph (\eg, web tables, spreadsheet). 
And more than 90\% of numerical fields can not match entity or property in KG (we perform statistics on non-WikiTable (Web table and spreadsheet) mentioned in \refsec{sec:corpus} and match with MTab\cite{Nguyen2019MTabMT}). For example, ``Final Exam'' in \reftab{tab:A} can not match KG, while it has ``Score'' measure type.

When determining the measure type taxonomy, we also collect manual labels on 882 tables (consisting of 6,715 fields) randomly sampled from our spreadsheet dataset. 
We start with a longer list of measure types as discussed in Definition\ref{eg:msr-type}, map DBPedia properties to the list for T2D, map Wikidata properties for SemTab and mark all 3,139 measures with types in the 882 sampled tables. As mentioned in \refsec{sec:msr-pair-type}, we only keep measure types with $\geq$10 labels (in T2D and sampled tables) or $\geq$100, resulting in 36,859 fields fall in our measure taxonomy in \reftab{tab:msr-taxonomy}.
%, covering \TODO{90.20\%} of all the measures we see in the merged dataset. 
Although TURL dataset also has property (relation extraction in its paper) labels, there are less than 1\% labeled as measure property, so they are not used in this task.

\subsection{Default Aggregation Function}
An essential operation in data analysis is to aggregate multiple measurements from a field, usually grouped by another  breakdown dimension. 
For each measure, there are some AGG functions more widely applied to it. The most suitable AGG function could be adopted as default calculation by downstream analysis tasks. For most measures, AVG can be applied directly, but often SUM is a better choice if possible. For some downstream scenarios, AVG/SUM can cover most usage cases.

\begin{figure}[htbp]
    \centering
    \includegraphics[width=.75\columnwidth]{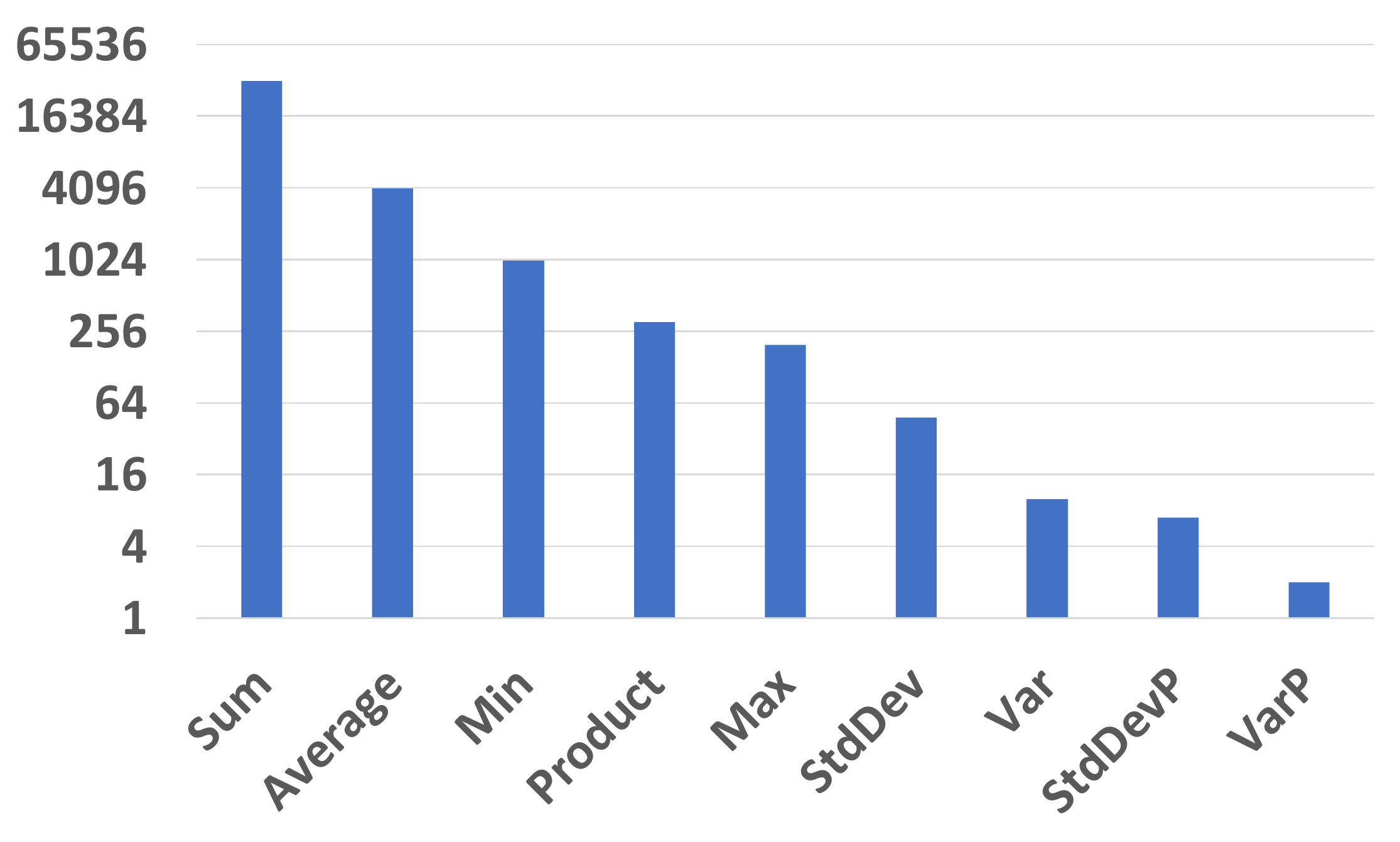}
    \caption{Default Aggregation Functions. (Taking the logarithm of y-axis with base 4)}
    \label{fig:agg_funcs}
    \vspace{-3mm}
\end{figure}

\subsection{Problem Formulation}
\label{sec:app-formulation}
The machine learning formulation of the metadata is a Binary classification of measure/dimension for each field of a given table. 

To help downstream tasks on prioritizing search order and re-ranking results, we formulate common field roles as three machine learning tasks providing 0$\sim$1 commonness score (higher means more preferred in general) for each field of a given table and recommend fields list for each table in order of commonness score.

The machine learning task for measure pair identification is a binary classification of any pair of numerical fields within a given table. For measure type and dimension type, it is a 19-way and 255-way classification problem.

The machine learning task for the default aggregation function is to provide 0$\sim$1 ranking scores for popular AGG functions. 
\section{Corpus}

\subsection{Dataset Preprocessing}

To avoid data leakage and imbalance, we carry out the following steps. Note that for a fair comparison, we adopt TURL dataset train/valid/test split, so do not perform the following steps. 
\begin{enumerate}
    \item \textit{Table Deduplication}. To avoid the ``data leakage'' problem that duplicated tables are allocated into both training and testing sets, tables are grouped according to their schemas\footnote{Two tables are defined to have the same \textbf{schema} if they have the same number of fields, and each field's data type and header name are correspondingly equal.}. 
    \item \textit{Down Sampling}. After deduplication, the number of tables within each schema is very imbalanced -- 0.23\% schemas cover 20\% of these tables. To mitigate this problem, we randomly sample unique tables under the threshold (11 for the Chart dataset, 2 for the Pivot dataset, 1 for Vendor \& T2D \& Semtab dataset).
\end{enumerate}

The schemas are randomly allocated for training, validation, and testing in the ratio of 7:1:2. 

For spreadsheet datasets, we follow the steps in \cite{zhou2021charts} and \cite{Zhou2020Table2AnalysisMA}, including extraction charts and pivot tables. 
% For TURL datasets, we only keep tables with column type in \cite{deng2020TURL}. Because they only keep column type field header in 

For all six datasets, we extract data features (\refsec{sec:data_features}) and map knowledge graph (\refsec{sec:knowledge_fusion}) as the input of our models.

% \subsection{Knowledge Graph Matching}
% \label{sec:app_kg}

\subsection{Data Quality}
\label{sec:app-quality}

The inspection focused on the Chart and Pivot datasets, as the quality of the T2D, TURL, and SemTab datasets had already been checked by their authors (for tasks with different labels, we also performed a manual mapping). We conducted the quality inspection with 5 experts who have analysis experience. All experts are from China. During the inspection, we randomly selected 100 tables from the Chart and Pivot datasets, and asked experts to score the labels of the corresponding tasks, and the score is between 0 and 1.  Our results showed that measure / dimension dichotomy got 0.99, common field roles got 0.97, aggregation functions got 0.93 and measure pair got 0.97 on average. This demonstrates that our corpus contains high-quality data and supervision.

\section{KDF Framework}
\begin{figure*}
    \centering
   \includegraphics[width=2 \columnwidth]{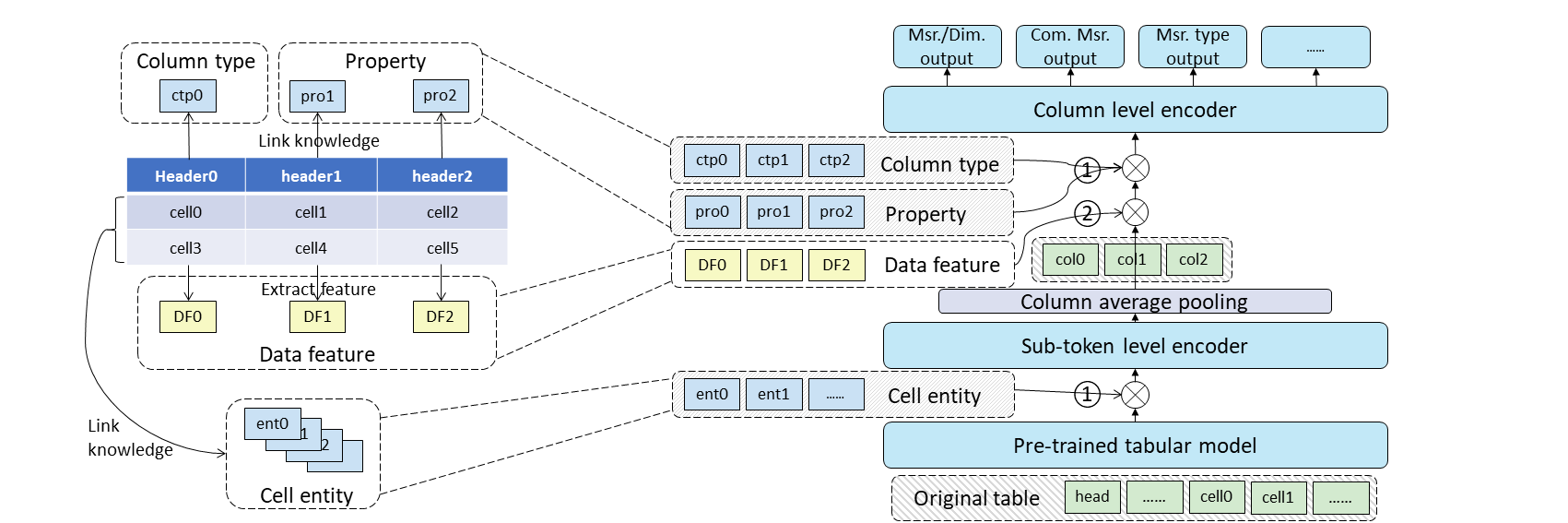}
    \caption{Overall metadata model. \Circled{1} represents knowledge fusion module in \reffig{fig:knowledge_fusion}, and \Circled{2} represents distribution fusion module in \reffig{fig:distribution_fusion}.}
    \label{fig:model_graph}
    \vspace{-3mm}
\end{figure*}
\subsection{Overall Model Architecture}
\label{sec:app-overall}
KDF overall architecture is as follows:

(1) Pre-trained tabular models are used as preliminary encoders generating initial representation for table elements. After the preliminary encoding phase(\textbf{``Pre-trained tabular model''} in \reffig{fig:model_graph}), we get a sub-token level or cell level (according to different pre-trained tabular models) table embedding sequence. 

(2) We fuse knowledge of cell entity and tabular model embedding with \textbf{``Knowledge fusion''}, and pass it into sub-token (or cell) level Transformer\cite{attention_is_all_you_need} encoder(\textbf{``Sub-token level encoder''}). 

(3) We apply \textbf{average pooling} to get an embedding representation for each column. For each column, we use \textbf{``Distribution fusion''} to fuse distribution from data features and \textbf{``Knowledge fusion''} to fuse knowledge of column type and property in order. Then we pass those column embeddings into Transformer encoder (\textbf{``Column level encoder''}) and linear output heads for each metadata task. 

\subsection{Knowledge Fusion Module}
For both entity and property, existing knowledge representation covers comprehensively and performs well. However, if their embedding is trained with table corpus, it can only cover limited entity and property. To increase the extensibility and performance of the model, we directly use knowledge representation in OpenKE\footnote{ http://openke.thunlp.org/}.

There are three kinds of knowledge information linked with knowledge graph according to table interpretation tasks -- cell entity, column type, and property. For a table with knowledge linking (\eg TURL dataset, Semtab dataset), we utilize original knowledge linking. Otherwise, we adopt MTab\footnote{https://github.com/phucty/mtab\_tool} to link the knowledge graph. MTab is a tool to annotate tables with knowledge graphs, and they get first place in Semtab2019\cite{jimenez2019semtab} and Semtab2020\cite{cutrona2020semtab}.

\subsection{Distribution Fusion Module}
\begin{figure}[tbp]
    \centering
    \includegraphics[width=0.9\columnwidth]{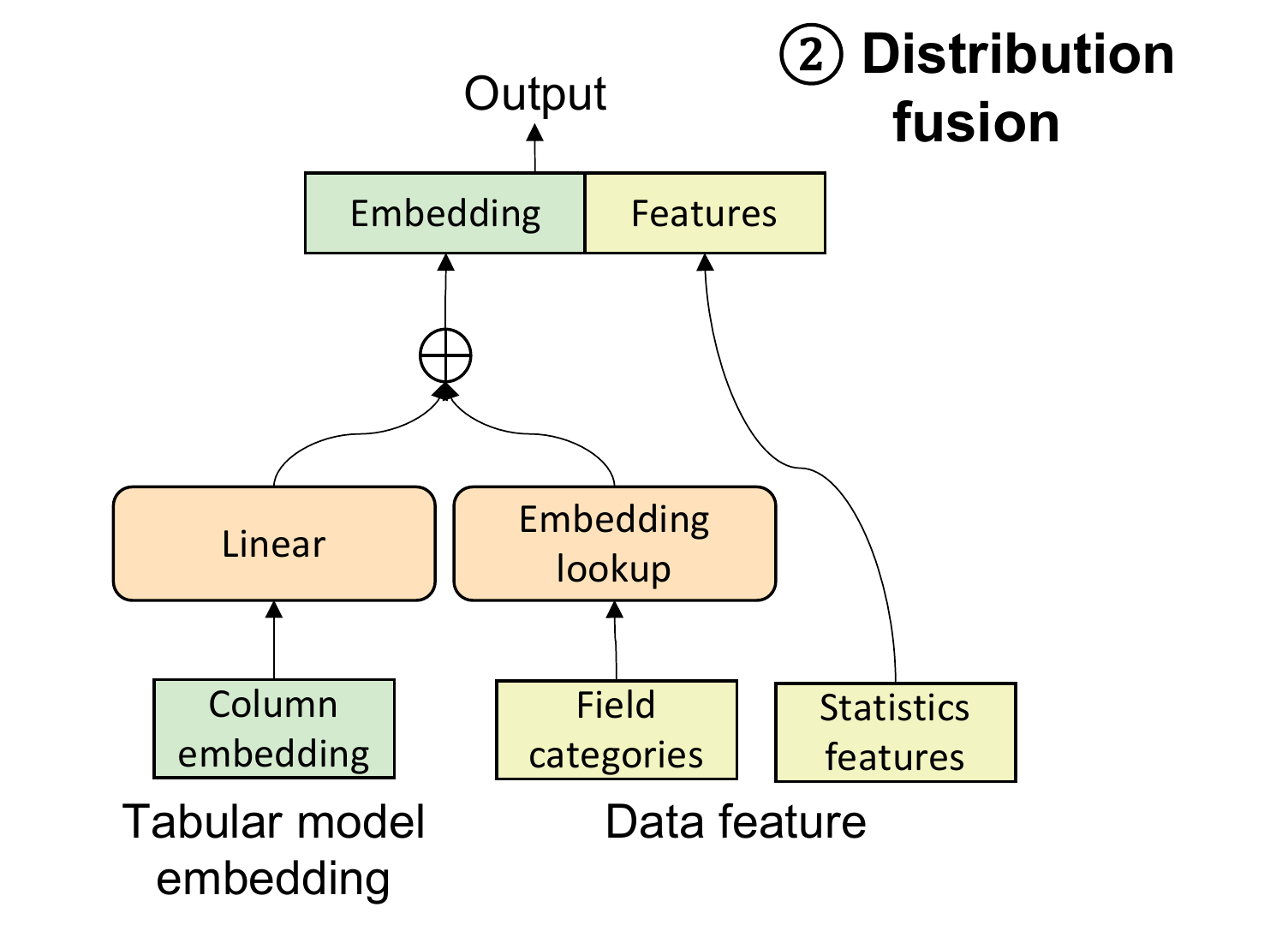}
	\caption{Distribution Fusion Module in KDF Framework. The module represents each token.}
    \label{fig:distribution_fusion}
	\vspace{-2mm}
\end{figure}
\label{sec:app-df}
(1) Statistics features: 

\quad $\bullet$ Progression features: ChangeRate, PartialOrdered, OrderedConfidence, ArithmeticProgressionConfidence, GeometricProgressionConfidence.
        
\quad $\bullet$ String features: AggrPercentFormatted, medianLen, LengthStdDev, AvgLogLength, CommonPrefix, CommonSuffix, Cardinality, AbsoluteCardinality.
        
\quad $\bullet$ Number range features: Aggr01Ranged, Aggr0100Ranged, AggrInteger, AggrNegative, SumIn01, SumIn0100.
        
\quad $\bullet$ Distribution features: Benford, Range, NumRows, KeyEntropy, CharEntropy, Variance, Cov, Spread, Major, Skewness, Kurtosis, Gini.

(2) Field categories: 
    
Including FieldType (Unknown, String, Year, DateTime, Decimal), IsPercent, IsCurrency, HasYear, HasMonth and HasDay.

\subsection{Multi-task Learning}
\label{sec:mtl}
We use 19-class Cross Entropy loss for measure type, and standard binary Cross Entropy loss for measure/dimension dichotomy, common measures and dimensions, and measure pair tasks. In common measures and dimensions task, we use scores of ``true'' class as their commonness scores. For aggregation function and dimension type task, we respectively use 7- and 255-class Cross Entropy loss and average loss of each field by the number of its ground truth, because there may be more than one ground truth for one field, which will lead to an unbalanced loss without averaging the fields.

\section{Experiment Details}
\subsection{Experiment Setting}

All the experiments are run on Linux machines with 24 CPUs, 448 GB memory, and 4 NVIDIA Tesla V100 16G-memory GPUs. We use hit rate ($HR@k=\frac{\#hits@k}{\#samples}$) to evaluate recommendation tasks on each table and accuracy to evaluate classification tasks on each field.

It's worth noting that to avoid label leaking, we do not use Knowledge fusion for column type and property in dimension type task. For the measure type task, we focus on the columns that can not be directly linked with properties from the knowledge graph for these columns are hard for existing table interpretation methods. To imitate those tables, we mask knowledge graph information for the field to train and evaluate in measure type task.

\subsection{Metadata Model Details}
\label{sec:app-KDF-exp}
In detail, we have the following model size:
\begin{itemize}
    \item Sub-token level: transformer encoder: 2 layers, 8 heads, dimension of embedding: $d_{tok}$ = 192, $d_{ent}$ = 100, $d_{h}$ = 64
    \item Column level: transformer encoder: 2 layers, 8 heads, dimension of embedding: $d_{tok}$ = 128, $d_{ent}$ = 100, $d_{h}$ = 64
\end{itemize}

We train Metadata on 10 epochs, 64 batch sizes, AdamW optimizer with $1\times 10^{-4}$ learning rate and 0.001 weight decay. We choose half visible hyperparameter $m=0.5$. It's worth noting that, due to insufficient supervision, aggregation function tasks use the result of epoch5.

\subsection{Benchmarks}
\label{sec:app-baseline}

\subsubsection{Rule-based}
\label{sec:app-rule}
The rule-based model predicts an output for every field based on the field surface attributes since it is off-the-shelf ready to use and does not involve additional training.
The specific rules for each metadata task are as follows:

(1) Measure/dimension dichotomy: The model predicts Measure when the input field is numerical (\ie, the field consists purely of numbers); otherwise; the model predicts Dimension.

(2) Nature key: The model predicts a field to be Natural Key if and only if the field is the leftmost field among fields with cardinality of 1 (\ie, the field contains all unique data values). 

(3) Common breakdown: The model predicts a field to be Common Breakdown if and only if the field is the leftmost among dimension fields whose cardinality is less than 0.4. 

(4) Common measures: The model predicts a field to be Common Measure if and only if the field is the rightmost numerical field.

(5) Dimension type: The model predicts every field to be of the type with most samples (sports.sports\_team in our case).

(6) Measure type: The model predicts every field to be of the type with the most samples (Count in our case).

(7) Measure pair: The model predicts two fields to be a Measure Pair if and only if they are contiguous Measure fields.

(8) Default Aggregation Function: The model predicts the default aggregation function of every field to be the function with most samples (SUM in our case).

\subsubsection{Traditional Machine Learning}
\label{sec:app-tradition}
Traditional machine learning is widely used due to both its effectiveness and efficiency. In data profiling\cite{abedjan2018dataprofiling}, there exist several state-of-art works that apply traditional learning with specifically extracted features. To compare with them, we design a traditional machine learning baseline by adopting a series of manually designed field data features (elaborated in \refsec{sec:data_features}). 
Traditional machine learning baseline implementation details of each metadata task are as follows:

(1) In binary classification tasks (\ie, measure/dimension dichotomy, measure pair), a single classification model is trained, and the model hyperparameters are determined based on the validation set metrics.

(2) In ranking tasks(\ie, common measures, common breakdown, and primary key), we train a binary classifier to make predictions for each field and rank the predictions with the raw probabilities given by the model outputs.

(3) In multi-class classification tasks (\ie dimension type, measure type, and default aggregation function), we directly train a multi-class classifier using traditional machine learning algorithms.

In the experiment, we choose GBDT,  Random Forest, Adaboost, and Naive Bayes to evaluation on each task, and display the performance of the best two baselines (GBDT and Random Forest) to compare with Metadata model. 

\subsubsection{Pre-trained Tabular Model}
\label{sec:app-turl}
\label{sec:TURL}
TURL is a structure-aware Transformer encoder to model the structural information about tables on table interpretation tasks and achieves Sate-Of-Art results. They also use Masked Entity Recovery pre-training objective to learn knowledge information about entities in relation tables. Since this model considers entity information, we select it as a strong baseline to show the effectiveness of Metadata model architecture in fusing the knowledge graph information. 
% The implementation details are shown in \refsec{sec:app-turl}.

To evaluate the performance of TURL in metadata tasks and have a fair comparison, we take the following steps:

(1) Map knowledge entity. Because entity linking in Semtab2020 dataset only has Wikidata id but TURL uses DBPedia id in embedding, we use WikiMapper\footnote{https://github.com/jcklie/wikimapper} to map Wikidata id to DBPedia id. It's worth noting that dimension type task is one of TURL's original tasks, so WikiMapper is not used in this task. Thus, this task can prove that Metadata still exceeds TURL without the influence of WikiMapper (details are shown in \refsec{sec:main_results}).

(2) Interpret the original table. We use TURL pre-trained encoder\footnote{https://github.com/sunlab-osu/TURL} to interpret the original table and get the embedding representation of flat tables.

(3) Train and evaluate metadata tasks. To do classification on metadata tasks, we use the same sub-token level encoder, column level encoder and loss function as our Metadata model while adopting the same training strategy with TURL.

For TAPAS and TABBIE, we  use the same sub-token level encoder, column level encoder in \reftab{fig:model_graph} (without fusion modules).

\subsubsection{Large Language Model \label{sec:app-llm}}
LLMs have recently gained significant attention for their success in various downstream tasks. In this study, we conduct a preliminary exploration of LLMs performance on metadata tasks. 
% The experimental setup is described in \refsec{sec:app-llm}. 
% \textcolor{red}{TODO: more results}.

\paragraph{Hyperparameter Setting} We set the decoding temperature to 0 (i.e., greedy decoding without sampling) since the question of metadata tasks has a unique answer. We set the max output length to 200.
\paragraph{Models} We select GPT-3 (text-davinci-001) and GPT-3.5 (text-davinci-002 and text-davinci-003) to conduct experiments since these models are shown to be powerful in many downstream tasks.

\paragraph{Zero / Few-shot Learning} We apply zero-shot and few-shot learning to evaluate the performance of large language models on metadata tasks.
For each setting, we add the definition of metadata terminology to allow LLMs for in-context learning.
Besides, the few-shot example is uniformly sampled from the training set for each task. 
Only high-quality examples with a non-empty header are selected.

\paragraph{Prompts}
We have created specific prompts for various tasks, which are listed at the end of this section. The optional `<example></example>` block includes a few-shot example.
Additionally, for the dimension type task with a large number of hierarchical choices, we employ a hierarchical questioning approach to avoid truncated input. 
For instance, we furnish the first-layer options (e.g. people, location, organization...) to the LLMs and the second-layer options (e.g. sports\_team, pro\_athlete, sports\_league...) to those questions answered with the correct first-layer choices. 
\begin{lstlisting}
**Task: measure / dimension**
<example>
Given the markdown table:
[linearized markdown table row by row]
Is the [ordinal number] column ("[header name]" column)  a measure or dimension? The answer is: [answer]
</example> (optional)
Given the markdown table:
[linearized markdown table row by row]
Is the [ordinal number] column ("[header name]" column) measure or dimension? ([term definition]) Please answer concisely a 'measure' or 'dimension'. (Do not return any explanation or any additional information.)
=>

**Task: natural key, common breakdown, common measure**
<example>
Given the markdown table:
[linearized markdown table row by row]
Which column is the natural key / common breakdown / common measure? The answer is: [answer]
</example> (optional)
Given the markdown table:
[linearized markdown table row by row]
Which column is the natural key / common breakdown / common measure with the highest probability? ([term definition]) Please answer a tuple of '(ordinal English word, header name)', where 'ordinal English word' starts from 'first'. (Do not return any explanation or any additional information.)
=>

**Task: dimension type, measure type, aggregation**
<example>
Given the markdown table:
[linearized markdown table row by row]
Which dimension type / measure type / aggregation is the [ordinal number] column ("[header name]" column)? The answer is: [answer]
</example> (optional)
Given the markdown table:
[linearized markdown table row by row]
Which of the following dimension types / measure types / aggregations is the [ordinal number] column ("[header name]" column)? (Do not return any explanation or any additional information.)
[choice 1]
[choice 2]
...
=>

**Task: measure pair**
<example>
Given the markdown table:
[linearized markdown table row by row]
Is the [ordinal number] column ("[header name]" column) and the [ordinal number] column ("[header name]" column) a measure pair? The answer is: [answer]
</example> (optional)
Given the markdown table:
[linearized markdown table row by row]
Is the [ordinal number] column ("[header name]" column) and the [ordinal number] column ("[header name]" column) a measure pair? [term definition] Please answer concisely a 'yes' or 'no'. (Do not return any explanation or any additional information.)
=>
\end{lstlisting}

\subsection{Results of Large Language Models}
In \reftab{tab:llm}, there are more results on large language models for metadata tasks.
We only report the results without terminology definition since we observe no significant improvement but degraded performance on measure / dimension dichotomy (-1\% on 0-shot, -5\% on 1-shot) and natural key (-4\% on 0-shot, -1\% on 1-shot) tasks on text-davinci-003.
% Table generated by Excel2LaTeX from sheet 'Sheet1'
\begin{table*}[htbp]
  \centering
  \caption{Large Language Model More Results on Metadata. ``text-davinci-001'' is GPT-3, and ``text-davinci-002'' and ``text-davinci-003'' are GPT-3.5.}
  \resizebox{\textwidth}{!}{
    \begin{tabular}{c|c|c|c|c|c|c|c|c}
    \toprule
    \multicolumn{1}{c|}{\multirow{2}[3]{*}{Model}} & Msr & Natural Key & Com. Breakdown & Com. Msr & Dim Type & Msr Type & Msr Pair & Aggregation \\
\cmidrule{2-9}          & Acc. & HR@1 & HR@1 & HR@1 & Acc. & Acc. & Acc. & Acc. \\
\midrule
    text-davinci-001 0-shot & 28.28\% & 73.66\% & 38.79\% & 26.03\% & 4.43\% & 13.10\% & 55.71\% & 19.49\% \\
    text-davinci-001 1-shot & 52.86\% & 78.01\% & 38.48\% & 22.85\% & 2.59\% & 10.92\% & 54.11\% & 35.84\% \\
    text-davinci-002 0-shot & 64.57\% & 88.10\% & 44.23\% & 43.08\% & 31.85\% & 18.64\% & 58.41\% & 65.26\% \\
    text-davinci-002 1-shot & 79.41\% & 74.03\% & 40.81\% & 57.36\% & 35.94\% & 16.37\% & 55.63\% & 55.01\% \\
    text-davinci-003 0-shot & 87.18\% & 88.39\% & 41.51\% & 50.18\% & 46.11\% & 35.10\% & 51.45\% & 77.46\% \\
    text-davinci-003 1-shot & 91.14\% & 82.47\% & 44.80\% & 57.30\% & 46.38\% & 29.32\% & 55.36\% & 75.66\% \\
    text-davinci-003 3-shot & 90.53\% & 78.81\% & 45.12\% & 54.14\% & 41.12\% & 28.43\% & 56.81\% & 80.37\% \\
    text-davinci-003 5-shot & 90.71\% & 79.23\% & 44.82\% & 56.44\% & 40.48\% & 29.18\% & 56.96\% & 80.47\% \\
    \bottomrule
    \end{tabular}}%
  \label{tab:llm}%
\end{table*}%

\subsection{Measure Type}
As mentioned in \refsec{sec:data_profiling}, there are several existing works focusing on the similar or same task. To compare dimension type, we apply the State-Of-Art table interpretation model -- TURL as a baseline as discussed in \refsec{sec:TURL}. To compare measure type, we apply the most relevant unit detection work \cite{yi2018recognizing} (RQN) as a baseline. It proposes a feature-based method to automatically determine the quantity names for column values. It uses hand-designed rules to extract features from raw tables in both column values and name. The extracted features are used to train a random forest classifier. The results are shown in \reftab{tab:msr_type_exp}.

\begin{table}%[htbp]
  \centering
  \caption{Measure Type Additional Results.}
  \resizebox{.7\linewidth}{!}{%
    \begin{tabular}{cc}
    \toprule
    Model & Measure Type Acc. \\
    \midrule
    RQN   & 65.55\% \\
    RF    & 70.09\% \\
    KDF+TAPAS w/o DF & \textbf{81.51\%} \\
    KDF+TABBIE w/o DF & 80.29\% \\
    \bottomrule
    \end{tabular}%
  }
  \label{tab:msr_type_exp}%
  \vspace{-3mm}
\end{table}%
 In \reftab{tab:msr_type_exp}, Metadata models outperform RQN more than 10\%. RQN has a limit to identifying quantities from the existing unit in the column, while Metadata model learns more.

\subsection{Ablation Results}
% Table generated by Excel2LaTeX from sheet 'Sheet1'
\begin{table*}[htbp]
  \centering
  \small
\centering
\caption{Ablation Results. Setting of this table is the same as \reftab{tab:exp_result}. ``DF'' -- Distribution fusion, ``KG'' -- Knowledge fusion. $\Delta$ means non-ablation models (KDF+TAPAS or KDF+TABBIE) metric minus corresponding evaluation metric.}
\label{tab:ablation_study}
\resizebox{\textwidth}{!}{
    \begin{tabular}{l|cc|cccccc|cccccc|cc}
    \toprule
    \multicolumn{1}{c|}{\multirow{2}[4]{*}{Model}} & \multicolumn{2}{c|}{Msr } & \multicolumn{2}{c}{Natural Key} & \multicolumn{2}{c}{Com. Breakdown} & \multicolumn{2}{c|}{Com. Measure} & \multicolumn{2}{c}{Dim Type } & \multicolumn{2}{c}{Msr Type} & \multicolumn{2}{c|}{Msr Pair} & \multicolumn{2}{c}{Aggregation} \\
\cmidrule{2-17}          & \multicolumn{1}{c}{Acc.} & \multicolumn{1}{c|}{$\Delta$} & \multicolumn{1}{c}{HR@1} & \multicolumn{1}{c}{$\Delta$} & \multicolumn{1}{c}{HR@1} & \multicolumn{1}{c}{$\Delta$} & \multicolumn{1}{c}{HR@1} & \multicolumn{1}{c|}{$\Delta$} & \multicolumn{1}{c}{Acc.} & \multicolumn{1}{c}{$\Delta$} & \multicolumn{1}{c}{Acc.} & \multicolumn{1}{c}{$\Delta$} & \multicolumn{1}{c}{Acc.} & \multicolumn{1}{c|}{$\Delta$} & \multicolumn{1}{c}{Acc.} & \multicolumn{1}{c}{$\Delta$} \\
    \midrule
    KDF+TAPAS & 98.45  & 0.00  & \textbf{95.04 } & 0.00  & 64.03  & 0.00  & 71.09  & 0.00  & \textbf{97.97 } & 0.00  & 79.32  & 0.00  & 77.50  & 0.00  & 91.48  & 0.00  \\
     \quad w/o DF & 97.40  & \cellcolor[rgb]{ .988,  .792,  .796}-1.05  & 94.01  & \cellcolor[rgb]{ .988,  .796,  .8}-1.03  & 61.66  & \cellcolor[rgb]{ .976,  .533,  .541}-2.36  & 70.87  & \cellcolor[rgb]{ .996,  .953,  .953}-0.22  & 97.93  & \cellcolor[rgb]{ .996,  .988,  .988}-0.04  & \textbf{81.51 } & \cellcolor[rgb]{ .529,  .667,  .839}2.19  & 75.82  & \cellcolor[rgb]{ .984,  .671,  .675}-1.68  & 89.76  & \cellcolor[rgb]{ .98,  .659,  .663}-1.72  \\
     \quad w/o KG & 98.54  & \cellcolor[rgb]{ .984,  .988,  .996}0.09  & 94.29  & \cellcolor[rgb]{ .992,  .851,  .851}-0.75  & 62.82  & \cellcolor[rgb]{ .988,  .761,  .765}-1.20  & 70.06  & \cellcolor[rgb]{ .988,  .796,  .796}-1.03  & 96.30  & \cellcolor[rgb]{ .984,  .671,  .675}-1.67  & 76.77  & \cellcolor[rgb]{ .976,  .498,  .502}-2.56  & 77.39  & \cellcolor[rgb]{ .996,  .976,  .976}-0.10  & \textbf{91.51 } & \cellcolor[rgb]{ .996,  .996,  1}0.03  \\
    \quad  w/o DF KG & 97.09  & \cellcolor[rgb]{ .984,  .733,  .733}-1.36  & 93.79  & \cellcolor[rgb]{ .988,  .753,  .757}-1.25  & 58.62  & \cellcolor[rgb]{ .973,  .412,  .42}-5.40  & 70.01  & \cellcolor[rgb]{ .988,  .784,  .788}-1.09  & 96.21  & \cellcolor[rgb]{ .98,  .651,  .659}-1.76  & 78.49  & \cellcolor[rgb]{ .992,  .835,  .839}-0.83  & 75.20  & \cellcolor[rgb]{ .976,  .549,  .553}-2.29  & 89.96  & \cellcolor[rgb]{ .984,  .702,  .706}-1.52  \\
\midrule
KDF+TABBIE & \textbf{98.93 } & 0.00  & 94.99  & 0.00  & \textbf{65.49 } & 0.00  & \textbf{71.74 } & 0.00  & 97.33  & 0.00  & 78.74  & 0.00  & \textbf{79.01 } & 0.00  & 90.58  & 0.00  \\
     \quad w/o DF & 97.89  & \cellcolor[rgb]{ .988,  .792,  .796}-1.04  & 94.78  & \cellcolor[rgb]{ .996,  .957,  .957}-0.21  & 62.64  & \cellcolor[rgb]{ .973,  .439,  .447}-2.85  & 71.60  & \cellcolor[rgb]{ .996,  .969,  .973}-0.14  & 97.34  & 0.01  & 80.29  & \cellcolor[rgb]{ .667,  .765,  .886}1.55  & 78.08  & \cellcolor[rgb]{ .988,  .816,  .816}-0.93  & 88.63  & \cellcolor[rgb]{ .98,  .616,  .624}-1.95  \\
     \quad w/o KG & 98.90  & \cellcolor[rgb]{ .996,  .992,  .992}-0.04  & 94.87  & \cellcolor[rgb]{ .996,  .976,  .976}-0.11  & 63.33  & \cellcolor[rgb]{ .976,  .573,  .58}-2.16  & 71.76  & \cellcolor[rgb]{ .996,  1,  1}0.02  & 95.38  & \cellcolor[rgb]{ .98,  .616,  .62}-1.96  & 76.35  & \cellcolor[rgb]{ .976,  .529,  .537}-2.39  & 77.99  & \cellcolor[rgb]{ .988,  .796,  .8}-1.02  & 90.71  & \cellcolor[rgb]{ .973,  .98,  .992}0.14  \\
     \quad w/o DF KG & 97.49  & \cellcolor[rgb]{ .984,  .714,  .718}-1.45  & 94.42  & \cellcolor[rgb]{ .992,  .886,  .89}-0.56  & 62.36  & \cellcolor[rgb]{ .973,  .412,  .42}-3.13  & 71.16  & \cellcolor[rgb]{ .992,  .886,  .886}-0.58  & 95.34  & \cellcolor[rgb]{ .98,  .608,  .612}-1.99  & 77.39  & \cellcolor[rgb]{ .984,  .733,  .737}-1.36  & 77.70  & \cellcolor[rgb]{ .984,  .741,  .741}-1.32  & 88.43  & \cellcolor[rgb]{ .976,  .576,  .584}-2.15  \\
\bottomrule

    \end{tabular}%
    }
    \vspace{-2mm}
\end{table*}%

% \subsubsection{Ablating DF and KG}
% \begin{figure}
%     \centering
%     \includegraphics[width=.75\columnwidth]{figures/gby_row_num.PNG}
%     \caption{Common Group By HR@1 by Row Number.}
%     \label{fig:gby_row_num}
%     \vspace{-3mm}
% \end{figure}

% \begin{figure}
%     \centering
%     \includegraphics[width=.75\columnwidth]{figures/msr_token_num.PNG}
%     \caption{Common measure HR@1 by token number. Token number is the length of the sequence after TAPAS tokenizer.}
%     \label{fig:msr_token_num}
%     \vspace{-5mm}
% \end{figure}

We ablate Distribution Fusion and Knowledge Fusion respectively and together, and the results are shown in \reftab{tab:ablation_study}. 

\label{sec:ablation}

Successful distribution and knowledge fusion gain the performance of models together. Compared with ablation models, for both TAPAS and TABBIE, the top-1 scores come most frequently from the non-ablation Metadata model. Especially, KG and DF boost 5.40 \%  (3.13 \%) for common breakdown task with Metadata TAPAS (TABBIE), 1.76\% (1.99\%) for dimension type, and 2.29\% (1.32\%) for measure pair. For example, on the measure pair task, all measure fields are numerical fields, so number understanding is the key point for this task. Data features fuse statistics information to help the model understand the full picture of numbers in one column. knowledge graph provides existing knowledge to understand each record and column and helps the model to understand the entire table, which is important for measure pair task.
%The model could get additional information from the knowledge graph to enhance the understanding of the table, at the same time, the field has a better distribution representation with Distribution fusion.

Especially on Common Goup By and Common Measure tasks, data statistics feature information is of great help to understand tables. For Common breakdown tasks, DF improves performance by 2.36\% (2.85\%) on Metadata TAPAS (TABBIE). Breakdown often appears in long tables (\eg average row number in the pivot dataset is 5805), due to the limitation of sequence length and comprehension, it's hard to capture the distribution of values for the pre-trained tabular model. Statistics feature is a good choice for understanding distribution. It is worth noting that DF does not benefit all tasks, \eg, measure type task. 

Especially on dimension type and measure type tasks, knowledge graph information is of great help to understand tables. For measure type task, KG improves performance by 3.02\% (2.90\%) on Metadata TAPAS (TABBIE) (w/o DF - w/o DF KG). The model could get additional information from the knowledge graph to enhance the understanding of the table. And it learns the useful pattern between measure type and cell entity/column type. 

The replaceable pre-trained tabular model brings a good opportunity to get better performance. KDF(TABBIE) outperforms KDF(TAPAS) on most tasks. Besides, KDF framework outperforms TURL on all tasks. In addition to distribution and knowledge infusing, the replaceable pre-trained tabular model is another key point. More and more pre-trained tabular models are emerging with better performance, and KDF model can be based on almost all transformer-based pre-training models. Thus, it's convenient to replace them with the models in Metadata, which can further improve performance and choose the best one.

\subsection{Downstream Interfaces}
\label{sec:downstream}
In our experiments on the Table2Charts model, we followed the same setup as described in \cite{zhou2021charts} and reported the results of the pretraining stage. In the metadata embeddings interface experiments, we used column embeddings from the KDF framework instead of FastText embeddings in the Table2Charts model. To ensure fair comparisons, we used TAPAS column embeddings as the input for the Table2Charts model, as the only difference between TAPAS column embeddings and KDF embeddings is the use of metadata knowledge. In the metadata IDs interface experiments, we used metadata ID tags on TAPAS column embeddings as the input for a fair comparison with the above experiments.

For TableQA tasks, we followed the T5-base experiment setup of UnifiedSKG~\cite{UnifiedSKG}. In the metadata sentences interface experiments, we first used our best KDF model to infer WikiTQ and HybridQA tables. Then, we concatenated metadata sentences, such as [Measure], [Natural Key], [Sum], etc., after the column headers. Finally, we fine-tuned the T5-base model using the same approach as in UnifiedSKG. In the metadata pretraining interface, we first continued pretraining the T5-base model using metadata tasks formulated as table question-answer tasks. The questions and answers were the same as the prompts described in \refsec{sec:app-llm}. Then, we fine-tuned the T5-base model using the same approach as in UnifiedSKG. We experimented with all combinations of four metadata tasks and reported the best results.

\end{document}